\pdfoutput=1  

\documentclass[12pt]{article}

\usepackage{tocloft}
\usepackage{graphicx}
\usepackage{natbib}
\usepackage[protrusion=true,expansion=true]{microtype}
\usepackage{amsmath,amsthm,amsfonts,amssymb,mathrsfs,dsfont,mathtools}
\usepackage{booktabs}
\usepackage{url}
\usepackage{float}
\usepackage{color}
\usepackage{parskip}
\usepackage[explicit]{titlesec}
\usepackage{threeparttable}
\usepackage{setspace}
\usepackage[margin=0.85in]{geometry}
\usepackage[font = onehalfspacing]{caption}
\usepackage[usenames,dvipsnames]{xcolor}
\usepackage[hyperfootnotes=true]{hyperref}
\usepackage[labelfont=bf]{caption}
\usepackage{authblk}
\usepackage{caption}
\usepackage[titletoc,toc,title]{appendix}
\usepackage{enumerate}
\usepackage{soul}
\usepackage{subfig}
\usepackage{subfiles} 
\usepackage[capitalise, nameinlink]{cleveref}
\usepackage{dsfont}
\usepackage{chngpage} 
\usepackage{algorithm}
\usepackage{algpseudocode}

\newcommand{\E}{\mathbb{E}}

\usepackage{scalerel,stackengine}
\stackMath
\newcommand\reallywidehat[1]{%
	\savestack{\tmpbox}{\stretchto{%
			\scaleto{%
				\scalerel*[\widthof{\ensuremath{#1}}]{\kern-.6pt\bigwedge\kern-.6pt}%
				{\rule[-\textheight/2]{1ex}{\textheight}}
			}{\textheight}%
		}{0.5ex}}%
	\stackon[1pt]{#1}{\tmpbox}%
}

\DeclareMathOperator*{\argmin}{arg\,min}

\DeclarePairedDelimiter{\ceil}{\lceil}{\rceil}

\usepackage{graphicx}
\newsavebox\CBox

\makeatletter
\newcommand*\bigcdot{\mathpalette\bigcdot@{.5}}
\newcommand*\bigcdot@[2]{\mathbin{\vcenter{\hbox{\scalebox{#2}{$\m@th#1\bullet$}}}}}
\makeatother

\setstcolor{red} 

\numberwithin{equation}{section}

\hypersetup{colorlinks = true, citecolor = MidnightBlue, linkcolor = MidnightBlue, urlcolor = MidnightBlue}

\titleformat{\section}{\normalfont\large\bfseries}{\thesection}{1em}{#1}
\titleformat{\subsection}{\normalfont\normalsize\bfseries}{\thesubsection}{1em}{#1}
\titleformat{\subsubsection}{\normalfont\normalsize\itshape}{\thesubsubsection}{1em}{#1}
\titlespacing\section{0pt}{12pt plus 4pt minus 2pt}{6pt plus 2pt minus 2pt}
\titlespacing\subsection{0pt}{12pt plus 4pt minus 2pt}{3pt plus 2pt minus 3pt}
\titlespacing\subsubsection{0pt}{12pt plus 4pt minus 2pt}{0pt plus 2pt minus 3pt}

\def\boxit#1{\vbox{\hrule\hbox{\vrule\kern6pt
			\vbox{\kern6pt#1\kern6pt}\kern6pt\vrule}\hrule}}

\definecolor{orange}{rgb}{1,0.5,0}
\definecolor{MyDarkBlue}{rgb}{0,0.08,0.45}

\newtheorem{remark}{Remark}[section]

\newtheorem{Def}{Definition}[section]

\begin{document}
	
	\title{\Large \bfseries Deep Equal Risk Pricing of Financial Derivatives with Multiple Hedging Instruments
	}
	
	\author[a] {Alexandre Carbonneau\thanks{Corresponding author\vspace{0.2em}. \newline
			{\it Email addresses:} \href{mailto:alexandre.carbonneau@mail.concordia.ca}{alexandre.carbonneau@mail.concordia.ca} (Alexandre Carbonneau), \href{mailto:frederic.godin@concordia.ca}{frederic.godin@concordia.ca} (Fr\'ed\'eric Godin).}}
	\author[b]{Fr\'ed\'eric Godin}
	\affil[a,b]{{\small Concordia University, Department of Mathematics and Statistics, Montr\'eal, Canada}}
	
	\vspace{-10pt}
	\date{ 
		\today}
	
	
	\maketitle \thispagestyle{empty} 
	
	
	\begin{abstract} 
		\vspace{-5pt}
		This paper studies the equal risk pricing (ERP) framework for the valuation of European financial derivatives. This option pricing approach is consistent with global trading strategies by setting the premium as the value such that the residual hedging risk of the long and short positions in the option are equal under optimal hedging. The ERP setup of \cite{marzban2020equal} is considered where residual hedging risk is quantified with convex risk measures. The main objective of this paper is to assess through extensive numerical experiments the impact of including options as hedging instruments within the ERP framework. The reinforcement learning procedure developed in \cite{carbonneau2020equal}, which relies on the deep hedging algorithm of \cite{buehler2019deep}, is applied to numerically solve the global hedging problems by representing trading policies with neural networks. Among other findings, numerical results indicate that in the presence of jump risk, hedging long-term puts with shorter-term options entails a significant decrease of both equal risk prices and market incompleteness as compared to trading only the stock. Monte Carlo experiments demonstrate the potential of ERP as a fair valuation approach providing prices consistent with observable market prices. Analyses exhibit the ability of ERP to span a large interval of prices through the choice of convex risk measures which is close to encompass the variance-optimal premium.
		
		\noindent \textbf{Keywords:} Equal risk pricing, Deep hedging, Convex risk measure, Reinforcement learning.
		
	\end{abstract} 
	\medskip

	\thispagestyle{empty} \vfill \pagebreak
	
	\setcounter{page}{1}
	\pagenumbering{roman}
	
	
	
	\doublespacing
	
	\setcounter{page}{1}
	\pagenumbering{arabic}

	\section{Introduction}
	In the famous setup of \cite{black1973pricing} and \cite{merton1973theory}, every contingent claim can be perfectly replicated through continuous trading in the underlying stock and a risk-free asset. These markets are said to be \textit{complete}, and derivatives are redundant securities with a unique arbitrage-free price equal to the initial value of the replicating portfolio. However, the gigantic size of the derivatives market demonstrates unequivalently that options are non-redundant and provide additional value above the exclusive trading of the underlying asset from the standpoint of speculation, risk management and arbitraging \citep{hull2003options}.
	Such value-added of derivatives in the real world stems from market incompleteness which
	arises from several stylized features of market dynamics such as discrete-time trading, equity risk (e.g. jump and volatility risks) and market impact (e.g. trading costs and imperfect liquidity). In contrast to the complete market paradigm, in incomplete markets, the price of a derivative cannot be uniquely specified by 
	a no-arbitrage argument since perfect replication is not always possible. 
	
	The problem of determining the value of a derivative is intrinsically intertwined with its corresponding hedging strategy. On the spectrum of derivative valuation procedures in incomplete markets, one extreme possibility is the so-called \textit{super-hedging strategy}, where the derivative premium is set as the value such that the residual hedging risk of the seller is nullified. However, the super-hedging premium is in general very large and is thus most often deemed impractical \citep{gushchin2002bounds}.
	On the other hand, a more reasonable and practical derivative premium entails that some level of risk cannot be hedged away and is thus intrinsic to the contingent claim. An additional layer of complexity to the hedging problem in incomplete markets is in selecting not only the sequence of investments in trading instruments, but also the \textit{category} of hedging instruments in the design of optimal hedges. Indeed, some categories of instruments are more effective to mitigate certain risk factors than others. For instance, it is well-known that in the presence of random jumps, option hedges are much more effective than trading exclusively the underlying stock due to the convex property of derivatives prices (see, for instance, \cite{coleman2007robustly} and \cite{carbonneau2020deep}). More generally, the use of option hedges dampens tail risk stemming from different risk factors (e.g. jump and volatility risks). The focus of this paper lies precisely on studying a derivative valuation approach called \textit{equal risk pricing} (ERP) for pricing 
	European derivatives consistently with optimal hedging strategies trading in various categories of hedging instruments (e.g. vanilla calls and puts as well as the underlying stock). 

	The ERP framework introduced by \cite{guo2017equal} determines the \textit{equal risk price} (i.e. the premium) of a financial derivative as the value such that the long and short positions in the contingent claim have the same residual hedging risk under optimal trading strategies. An important application of ERP in the latter paper is for pricing derivatives in the presence of short-selling restrictions for the underlying stock. Various studies have since extended this approach: \cite{ma2019pricing} provide Hamilton-Jacobi-Bellman equations for the optimization problem and establish additional analytical pricing formulas for equal risk prices, \cite{he2020revised} generalize the problem of pricing derivatives with short-selling restriction for the underlying by allowing for short trades in a correlated asset and \cite{alfeus2019empirical} perform an empirical study of equal risk prices when short selling is banned. One crucial pitfall of the \cite{guo2017equal} framework considered in all of the aforementioned papers is that the optimization problem required to be solved for the computation of equal risk prices is very complex. Consequently, closed-form solutions are restricted to very specific setups (e.g. Black-Scholes market) and no numerical scheme has been proposed to account for more realistic market assumptions. 
	
	\cite{marzban2020equal} recently extended the ERP framework by considering the use of convex risk measures under the physical measure to quantify residual hedging risk. A major benefit of the ERP setup of the latter paper is that it does not require the specification of an equivalent martingale measure (EMM), which is arbitrary in incomplete markets since there is an infinite set of EMMs \citep{harrison1981martingales}. Also, using convex measures to quantify residual risk is shown in \cite{marzban2020equal} to significantly reduce the complexity of computing equal risk prices; the optimization problem essentially boils down to solving two distinct non-quadratic global hedging problems, one for the long and one for the short position in the option. Dynamic programming equations are provided in \cite{marzban2020equal} for the aforementioned global hedging problems. However, it is well-known that traditional dynamics programming procedures are prone to the curse of dimensionality when the state and action spaces gets too large \citep{powell2009you}. The main objective of this current paper consists in studying the impact of trading different and possibly multiple hedging instruments 
	on the ERP framework, which thus necessitates large action spaces. Furthermore, a specific focus of this study is on assessing the interplay between different equity risk factors (e.g. jump and volatility risks) and the use of options as hedging instruments. Consequently,  
	large state spaces are also required to model the dynamics of the underlying stock and to characterize the physical measure dynamics of the implied volatility of options used as hedging instruments. A feasible numerical procedure in high-dimensional state and action spaces is therefore essential to this paper.
	
	\cite{carbonneau2020equal} expanded upon the work of \cite{marzban2020equal} by developing a tractable solution with reinforcement learning to compute equal risk prices in high-dimensional state and action spaces. The approach of the foremost study relies on the deep hedging algorithm of \cite{buehler2019deep} to represent the long and short optimal trading policy with two distinct neural networks. One of the most important benefits of parameterizing trading policies as neural networks is that the computational complexity increases marginally with the dimension of the state and action spaces. \cite{carbonneau2020equal} also introduce novel $\epsilon$-completeness metrics to quantify the level of market incompleteness which will be used throughout this current study. Several papers have studied different aspects of the class of deep hedging algorithms: \cite{buehler2019deep_2} extend upon the work of \cite{buehler2019deep} by hedging path-dependent contingent claims with neural networks, \cite{carbonneau2020deep} presents an extensive benchmarking of global policies parameterized with neural networks to mitigate the risk exposure of very long-term contingent claims, \cite{cao2020discrete} show that the deep hedging algorithm provides good approximations of optimal initial capital investments for variance-optimal hedging problems and \cite{horvath2021deep} deep hedge in a non-Markovian framework with rough volatility models for risky assets.
	
	The main objective of this paper consist in the assessment of the impact of using multiple hedging instruments on the ERP framework through exhaustive numerical experiments. To the best of the authors' knowledge, this is the first study within the ERP literature that considers trades involving options in the design of optimal hedges.
	The performance of these numerical experiments heavily relies on the use of reinforcement learning procedures to train neural networks representing trading policies and would be hardly reachable with other numerical methods. The first key contribution of this paper consists in providing a broad analysis of the impact of jump and volatility risks on equal risk prices and on our $\epsilon$-completeness metrics.
	These assessments expand upon the work of \cite{carbonneau2020equal} in two ways. First, the latter paper conducted sensitivity analyses of the ERP framework under different risky assets dynamics by trading exclusively with the underlying stock, not with options. However, the use of options as hedging instruments in the presence of such risk factors allows for the mitigation of some portion of unattainable residual risk when trading exclusively with the stock. Second, this current paper examines the sensitivity of equal risk prices and residual hedging risk to different levels of jump and volatility risks through a range of empirically plausible model parameters for asset prices dynamics (e.g. frequent small jumps and rare extreme jumps). The motivation is to provide new qualitative insights into the interrelation of different stylized features of jump and volatility risks on the ERP framework that are more extensive than in previous work studies. 
	The main conclusions of these experiments of pricing $1$-year European puts are summarized below. 
	\begin{itemize}
		\item [1)] In the presence of downward jump risk, numerical values indicate that hedging with options entails significant reduction of both equal risk prices and on the level of market incompleteness as compared to hedging solely with the underlying stock. The latter stems from the fact that while the residual hedging risk of both the long and short positions in $1$-year puts decreases when short-term option trades are used for mitigating the presence of jump risk, a larger decrease is observed for the short position due to jump risk dynamics entailing predominantly negative jumps. These results further demonstrate that options are non-redundant securities as it is the case in the Black-Scholes world.
		\item [2)] In the presence of volatility risk, numerical experiments demonstrate that while the use of options as hedging instruments can entail smaller derivative premiums, the impact can also be marginal and is highly sensitive to the moneyness level of the put option being priced as well as to the maturity of the traded options. This observation stems from the fact that contrarily to jump risk, volatility risk impacts both upside and downside risk. Thus, the use of option hedges does not necessarily benefit more the short position with a larger decrease of residual hedging risk as observed in the presence of jump risk. 
		\item [3)] The average price level of short-term options (i.e. average implied volatility level) used as hedging instruments is effectively reflected into the equal risk price of longer-term options. This demonstrates the potential of the ERP framework as a fair valuation approach consistent with observable market prices, which could be used, for instance, to price over-the-counter or long-term less liquid derivatives with short-term highly liquid options.
	\end{itemize}

	The last contribution of this paper is in benchmarking equal risk prices to derivative premiums obtained with \textit{variance-optimal hedging} \citep{schweizer1995variance}. Variance-optimal hedging procedures solve jointly for the initial capital investment and a self-financing strategy minimizing the expected squared hedging error. The optimized initial capital investment can be viewed as the production cost of the derivative, since the resulting dynamic trading strategy replicates the derivative's payoff as closely as possible in a quadratic sense.\footnote{
		Note that derivatives premiums prescribed by variance-optimal procedures coincide with the risk-neutral price obtained under the so-called \textit{variance-optimal martingale measure} \citep{schweizer1996approximation}. 
	} 
	The main motivation for these experiments is the popularity of variance-optimal hedging procedures in the literature for pricing derivatives. Furthermore, while these two derivative valuation procedures are both consistent with optimal trading criteria, the underlying global hedging problem of each approach treats hedging shortfall through a radically different scope. Indeed, equal risk prices obtained under the Conditional Value-at-Risk measure with large confidence level values, as considered in this paper, are the result of joint optimizations over hedging decisions to minimize tail risk of hedging shortfalls which penalize mainly (and most often exclusively) hedging losses, not gains. Conversely, variance-optimal procedures penalize equally hedging gains and losses, not solely losses. This benchmarking  of equal risk prices to variance-optimal premiums highlights the flexibility of ERP procedures for derivatives valuation through the choice of convex risk measure. Indeed, numerical values show that the range of equal risk prices obtained with several convex measures can be very large and is close to encompass the variance-optimal premium.

	The rest of the paper is as follows. \cref{section:equal_risk_pricing_framework} details the equal risk pricing framework considered in this study. \cref{section:neural_network} presents the numerical scheme to solve the optimization problem with the use of neural networks. \cref{section:numerical_results} performs extensive numerical experiments studying the equal risk pricing framework.
	\cref{section:conclusion} concludes. 
	

	\section{Equal risk pricing framework}
	\label{section:equal_risk_pricing_framework}
	This section details the equal risk pricing (ERP) framework considered in this paper, 
	which is an extension of the derivative valuation scheme introduced
	in \cite{marzban2020equal} 
	with the addition of multiple hedging instruments.  
	
	\subsection{Specification of the financial market}
	The financial market is in discrete-time with a finite time horizon of $T$ years and $N+1$ observation dates characterized by the set $\mathcal{T}:=\{t_n: t_n = n \Delta_N, n=0,\ldots,N\}$ where $\Delta_N:=T/N$. The probability space $(\Omega, \mathbb{P}, \mathcal{F})$ is equipped with the filtration $\mathbb{F}:=\{\mathcal{F}_n\}_{n=0}^{N}$ satisfying the usual conditions, where $\mathcal{F}_n$ contains all information available to market participants at time $t_n$. Assume $\mathcal{F}=\mathcal{F}_{N}$. $\mathbb{P}$ is referred to as the physical probability measure. 
	On each observation date, a total of $D+2$ financial securities can be traded on the market, which includes a risk-free asset, a non-dividend paying stock and $D$ standard European calls and puts on the latter stock whose maturity dates fall within $\mathcal{T}$. Let $\{B_{n}\}_{n=0}^{N}$ be the price process of the risk-free asset, where $B_{n}:=e^{r t_n}$ for $n=0,\ldots,N$ with $r \in \mathbb{R}$ being the annualized continuously compounded risk-free rate. 
	The definition of the price process for the risky securities 
	is now outlined. Since some of the tradable options can mature before the final time horizon $T$, the set of options that can be traded at the beginning of two different observation periods could differ. To reflect this modeling feature and properly represent gains of trading strategies, two different stochastic processes are defined, namely the price of tradable assets at the beginning and at the end of each period. First, let $\{\bar{S}_{n}^{(b)}\}_{n=0}^{N}$
	be the \textit{beginning-of-period risky price process} whose element $\bar{S}_{n}^{(b)}$ contains the time-$t_n$ price of all risky assets traded at time $t_n$. More precisely, $\bar{S}_{n}^{(b)}:=[S_{n}^{(0,b)}, \ldots, S_{n}^{(D,b)}]$ with $S_{n}^{(0,b)}$ and $S_{n}^{(j,b)}$ respectively being the time-$t_n$ price of the underlying stock and of the $j^{\text{th}}$ option that can be traded at time $t_n$ for $j=1,\ldots,D$. Similarly, let $\{\bar{S}_{n}^{(e)}\}_{n=0}^{N-1}$ be the \textit{end-of-period risky price process} where $\bar{S}_{n}^{(e)}:=[S_{n}^{(0,e)}, \ldots, S_{n}^{(D,e)}]$ with $S_{n}^{(0,e)}$ and $S_{n}^{(j,e)}$ respectively being the time $t_{n+1}$ price of the underlying stock and $j^{\text{th}}$ option that can be traded at time $t_n$. 
	Since the underlying asset is denoted as the risky asset with index $0$, $S_{n}^{(0,e)}=S_{n+1}^{(0,b)}$ for $n=0,\ldots,N-1$.
	Also, if the $j^{\text{th}}$ option that can be traded at $t_n$ matures at time $t_{n+1}$, then $S_n^{(j,e)}$ is the payoff of that option. In that case, $S_{n+1}^{(j,b)}$ is the price of a new contract with the same characteristics in terms of payoff function, moneyness level and time-to-maturity.
	Otherwise, $S_{n+1}^{(j,b)}= S_{n}^{(j,e)}$ holds for all time steps and all risky assets (i.e. for $j=0,\ldots,D$ and $n=0,\ldots,N-1$). An implicit assumption stemming from the latter equality is that trading in risky assets does not impact their prices. Moreover, for convenience, it is assumed throughout the current work that only options with a single-period time-to-maturity are traded, i.e. options are traded once and held until expiry.\footnote{
		Note that the optimization procedure for global policies described in \cref{section:neural_network} can naturally be generalized 
		for the case of rebalancing multiple times option contracts prior to their expiry.
	}
	
	This paper studies the problem of pricing a simple European-type derivative providing a time-$T$ payoff denoted by $\Phi(S_N^{(0,b)}) \geq 0$.\footnote{
		The derivative valuation approach presented in this paper can easily be adapted for European options whose payoff is of the form $\Phi(S_N^{(0,b)}, Z_N) \geq 0$ with $\{Z_n\}_{n=0}^{N}$ as some $\mathbb{F}$-adapted potentially multidimensional random process encompassing the path-dependence property of the payoff function. 
		For examples of such exotic derivatives, the reader is referred to \cite{carbonneau2020equal}.
	}
	For such purposes, the equal risk pricing scheme is considered, which entails optimizing two distinct self-financing dynamic trading strategies separately for both the long and short positions on the derivative, and then determining the premium which equates the residual hedging risk of the two hedged positions. 
	The mathematical formalism used for trading strategies in the current study is now outlined.
	A \textit{trading strategy}
	$\{\delta_{n}\}_{n=0}^{N}$ is an $\mathcal{F}$-predictable process\footnote{
	A process $X=\{X_{n}\}_{n=0}^{N}$ is said to be $\mathcal{F}$-predictable if $X_0$ is $\mathcal{F}_0$-measurable and $X_n$ is $\mathcal{F}_{n-1}$-measurable for $n=1,\ldots,N$.
	} 
	where $\delta_{n}:=[\delta_{n}^{(0)},\ldots,\delta_{n}^{(D)}, \delta_{n}^{(B)}]$ with $\delta_{n}^{(B)}$ and $\delta_{n}^{(j)}$, $j=0,\ldots,D$, respectively denoting the number of shares of the risk-free asset and the $j^{\text{th}}$ risky asset traded at time $t_{n-1}$ held in the hedging portfolio throughout the period $(t_{n-1},t_n]$, except for the case $n=0$ which represents the hedging portfolio composition exactly at time $t_0$. The notation $\delta_n^{(0:D)}:=[\delta_n^{(0)},\ldots,\delta_n^{(D)}]$ is used to define the vector containing exclusively positions in the risky assets. Furthermore, the initial capital investment of the trading strategy is always assumed to be completely invested in the risk-free asset, i.e. $\delta_0^{(B)}$ is the initial investment amount and $\delta_0^{(0:D)}:=[0,\ldots,0]$. 

	In this work, the trading strategies considered to hedge $\Phi$
	are obtained through a joint optimization over all trading decisions to minimize
	global risk exposure.
	Before formally describing the optimization problem, some well-known prerequisites from the mathematical finance literature are now provided; the reader is referred to \cite{lamberton2011introduction} for additional details. Let $\{V_{n}^{\delta}\}_{n=0}^{N}$ be the hedging portfolio value process associated with the trading strategy $\delta$, where $V_n^{\delta}$ is the time-$t_n$ portfolio value prior to rebalancing with $V_0^{\delta} := \delta_0^{(B)}$ and
	\begin{align}
		V_n^{\delta}:=\delta_n^{(0:D)} \bigcdot \bar{S}_{n-1}^{(e)} + \delta_n^{(B)}B_n, \quad n =1,\ldots, N,\label{eq:ref_hedging_port_val}
	\end{align}
	where $\bigcdot$ is the dot product operator.\footnote{
		For $X:=[X_1,\ldots,X_K]$ and $Y:=[Y_1,\ldots,Y_K]$, $X \bigcdot Y:=\sum_{j=1}^{K}X_j Y_j$.
	} Furthermore, denote as $\{G_{n}^{\delta}\}_{n=0}^{N}$ the discounted gain process associated with $\delta$ where $G_{n}^{\delta}$ is the time-$t_n$ discounted gain prior to rebalancing with $G_{0}^{\delta} := 0$ and 
	\begin{align}
		G_{n}^{\delta}:= \sum_{k=1}^{n} \delta_{k}^{(0:D)} \bigcdot (B_{k}^{-1} \bar{S}_{k-1}^{(e)} - B_{k-1}^{-1} \bar{S}_{k-1}^{(b)}), \quad n = 1,\ldots,N. \label{eq:ref_disc_gain_process_def}
	\end{align}
	The trading strategies considered in this paper are always \textit{self-financing}: they require no cash infusion nor withdrawal at intermediate times except possibly at the initialization of the strategy. More formally, a trading strategy is said to be self-financing if it is predictable 
	%
	%
	and if the following equality holds $\mathbb{P}$-a.s. for $n=0,\ldots,N-1$:
	\begin{align}
		\delta_{n+1}^{(0:D)} \bigcdot \bar{S}_n^{(b)} + \delta_{n+1}^{(B)}B_n = V_n^{\delta}. \label{eq:ref_self_financing}
	\end{align}
	Lastly, denote $\Pi$ as the set of accessible trading strategies, which includes all trading strategies that are self-financing and sufficiently well-behaved.
	\begin{remark}
		It can be shown that $\delta \in \Pi$ is self-financing if and only if $V_{n}^{\delta} = B_n(V_0^{\delta} + G_n^{\delta})$ holds $\mathbb{P}$-a.s. for $n=0,\ldots,N$; see for instance \cite{lamberton2011introduction}. The latter representation of portfolio values implies the following useful recursive equation \eqref{eq:ref_update_rule_port_value} to compute $V_{n}^{\delta}$ for $n=1,\ldots,N$ given $V_0^{\delta}$:
		\begin{align}
		V_{n}^{\delta} &= B_{n}(V_0^{\delta}+G_{n}^{\delta}) \nonumber
		\\ &= B_{n}(V_0^{\delta}+G_{n-1}^{\delta}+\delta_{n}^{(0:D)} \bigcdot (B_{n}^{-1} \bar{S}_{n-1}^{(e)} - B_{n-1}^{-1} \bar{S}_{n-1}^{(b)})) \nonumber
		\\ &= \frac{B_{n}}{B_{n-1}}V_{n-1}^{\delta} + \delta_{n}^{(0:D)} \bigcdot (\bar{S}_{n-1}^{(e)} - \frac{B_{n}}{B_{n-1}}\bar{S}_{n-1}^{(b)}) \nonumber
		\\ &= e^{r \Delta_N}V_{n-1}^{\delta} + \delta_{n}^{(0:D)}\bigcdot(\bar{S}_{n-1}^{(e)} - e^{r \Delta_N}\bar{S}_{n-1}^{(b)}). \label{eq:ref_update_rule_port_value}
		\end{align}
	\end{remark}
	
	\subsection{Equal risk pricing framework}
	\label{subsec:ERP_framework}	
	The financial market setting considered in this paper implies incompleteness stemming from discrete-time trading and equity risk factors (e.g. jump risk and volatility risk). 
	For the hedger, these many sources of incompleteness entail that most contingent claims are not attainable through dynamic hedging. Following the work of \cite{marzban2020equal} and \cite{carbonneau2020equal}, this study quantifies the level of residual hedging risk with convex risk measures as defined in \cite{follmer2002convex}. 
	
	\begin{Def}{(Convex risk measure)}
		For a set of random variables $\mathcal{X}$ representing liabilities and $X_1, X_2 \in \mathcal{X}$, $\rho : \mathcal{X} \rightarrow \mathbb{R}$ is a \textit{convex risk measure} if it satisfies the following properties:
		\begin{itemize}
			\item [0)] Normalized: $\rho(0) = 0$ (empty portfolio has no risk).
			\item [1)] Monotonicity: $X_1 \leq X_2 \Longrightarrow \rho(X_1) \leq \rho(X_2)$ (larger liability is riskier).
			\item [2)] Translation invariance: for $c \in \mathbb{R}$ and $X \in \mathcal{X}$, $\rho(X+c) = \rho(X) + c$ (borrowing amount $c$ increases the risk by that amount).  
			\item [3)] Convexity: for $c \in [0,1]$, $\rho(c X_1 + (1-c)X_2) \leq c \rho(X_1) + (1-c)\rho(X_2)$ (diversification does not increase risk).
		\end{itemize} 
	\end{Def}
	The hedging problem underlying the ERP framework is now formally defined.
	
	\begin{Def}{(Long- and short-sided risk)}
		\label{HedgingProbDef}
		For a given convex risk measure $\rho$, define $\epsilon^{(\mathcal{L})}(V_0)$ and $\epsilon^{(\mathcal{S})}(V_0)$ respectively as the measured risk exposure of a long and short position in $\Phi$ under the optimal hedge if the value of the initial hedging portfolio is $V_0 \in \mathbb{R}$:
		\begin{align}
			\epsilon^{(\mathcal{L})}(V_0) &:= \underset{\delta\in \Pi}{\min} \, \rho \left(-\Phi(S_{N}^{(0,b)}) -B_{N}(V_0 + G_{N}^{\delta})\right), \label{eq:risk_long}
			\\ \epsilon^{(\mathcal{S})}(V_0) &:= \underset{\delta \in \Pi}{\min} \, \rho \left(\Phi(S_{N}^{(0,b)}) - B_{N}(V_0 + G_{N}^{\delta})\right). \label{eq:risk_short}
		\end{align}
	\end{Def} 
	
	\begin{remark}
		\label{remark:inf_vs_min}
		As noted in \cite{carbonneau2020equal}, an assumption implicit to \cref{HedgingProbDef} is that the minimum in \eqref{eq:risk_long} or \eqref{eq:risk_short} is indeed attained by some trading strategy, i.e. that the infimum is in fact a minimum. 
	\end{remark}
	
	Note that the same risk measure $\rho$ is used for both the long and short positions global hedging problems. The rationale for this choice is threefold. First, considering the same convex risk measure for long and short positions
	is in line with the trading activities of some market participants that both buy and sell options with no directional view of the market. One example of such participant is a market maker of derivatives which typically expects to make a profit on bid-ask spreads, not by speculating \citep{basak2012dynamic}.
	Another motivation for using the same convex measure is for cases where a price quote must be given prior to knowing if the derivative is being purchased or sold. For instance, a client asks his broker to provide a quote for a derivative without revealing  his intention of buying or selling the option. A similar argument is made in \cite{bertsimas2001hedging} to motivate the use of a quadratic loss function for hedging shortfalls, which entails the same derivative price for the long and short position. Lastly, as shown in \cite{carbonneau2020equal}, using the same risk measure for both positions guarantees, under some specific conditions, that the ERP derivative premium is arbitrage-free.\footnote{
	Nevertheless, the authors want to emphasize that the numerical scheme developed in \cref{section:neural_network} for the global hedging problems \eqref{eq:risk_long} and \eqref{eq:risk_short} could easily be extended to include two distinct convex measures respectively for the long and short position hedges (see Remark $3.4$ of \cite{carbonneau2020equal} for additional details).
	} 

	It is interesting to note that the translation invariance property of $\rho$ entails that the optimal strategies solving \eqref{eq:risk_long}-\eqref{eq:risk_short}, denoted respectively by $\delta^{(\mathcal{L})}$ and $\delta^{(\mathcal{S})}$, are invariant to the initial capital investment amount $V_0$. The latter significantly enhances the tractability of the solution: 
		\begin{align}
			\delta^{(\mathcal{L})} &:= \underset{\delta\in \Pi}{\argmin} \, \rho \left(-\Phi(S_{N}^{(0,b)}) -B_{N}(V_0 + G_{N}^{\delta})\right) = \underset{\delta\in \Pi}{\argmin} \, \rho \left(-\Phi(S_{N}^{(0,b)}) -B_{N}G_{N}^{\delta}\right), \label{eq:hedging_risk_long}	
			\\ \delta^{(\mathcal{S})} &:= \underset{\delta \in \Pi}{\argmin} \, \rho \left(\Phi(S_{N}^{(0,b)}) - B_{N}(V_0 + G_{N}^{\delta})\right) = \underset{\delta \in \Pi}{\argmin} \, \rho \left(\Phi(S_{N}^{(0,b)}) - B_{N}G_{N}^{\delta}\right). \label{eq:ref_hedging_risk_short}
			%
		\end{align}
	Based on the aforementioned global hedging problems, the \textit{equal risk price} of a derivative is defined as the initial hedging portfolio value equating the measured risk exposures for both the long and short positions.
	
	\begin{Def}{(Equal risk price)}
		\label{def:ERP_option_price}
		The equal risk price $C_0^{\star}$ of $\Phi$ is defined as the real number $C_0$ such that
		\begin{align}
			\epsilon^{(\mathcal{L})}(-C_0) = \epsilon^{(\mathcal{S})}(C_0). \label{eq:ref_ERP_def}
		\end{align}
	\end{Def} 
	
	As shown for instance in \cite{marzban2020equal}, equal risk prices have the following representation
	which is used throughout the rest of the paper:
	\begin{align}
		C_{0}^{\star} = \frac{\epsilon^{(\mathcal{S})}(0) - \epsilon^{(\mathcal{L})}(0)}{2 B_{N}}. \label{eq:ref_equal_risk_prices}
	\end{align}	

	\cite{carbonneau2020equal} introduced the market incompleteness metric $\epsilon^{\star}$ defined as the level of residual risk faced by the hedgers of $\Phi$ if the hedged derivative price is set to $C_0^{\star}$ and optimal trading strategies are used by both the long and short position hedgers:\footnote{
	The last equality of \eqref{eq:ref_epsilon_def} can easily be obtained with the translation invariance property of $\rho$, see equation (8) of \cite{carbonneau2020equal} for the details.  
	}
	\begin{align}
		\epsilon^{\star}&:=\epsilon^{(\mathcal{L})}(-C_0^{\star}) = \epsilon^{(\mathcal{S})}(C_0^{\star}) = \frac{\epsilon^{(\mathcal{L})}(0) + \epsilon^{(\mathcal{S})}(0)}{2}. \label{eq:ref_epsilon_def}
	\end{align}
	
	%
	Consistently with the terminology of \cite{carbonneau2020equal}, $\epsilon^{\star}$ and $\epsilon^{\star}/C_0^{\star}$ are referred respectively as the measured residual risk exposure per derivative contract and per dollar invested. These $\epsilon^{\star}$-metrics will be extensively studied in numerical experiments conducted in \cref{section:numerical_results} to assess, for instance, the impact of the use of options as hedging instruments on the level of market incompleteness.
	
	
	\section{Deep equal risk pricing}
	\label{section:neural_network}
	The problem of solving the ERP framework, that is evaluating equal risk prices and $\epsilon$-completeness measures, boils down to the computation of the measured risk exposures $\epsilon^{(\mathcal{S})}(0)$ and $\epsilon^{(\mathcal{L})}(0)$. This section presents a reinforcement learning method to compute such quantities. The approach was first proposed in \cite{carbonneau2020equal} and relies on approximating optimal trading strategies with the \textit{deep hedging} algorithm of \cite{buehler2019deep} 
	through the representation of
	the long and short global trading policy with two distinct neural networks. In its essence, neural networks are a class of composite functions mapping \textit{feature vectors} (i.e. input vectors) to \textit{output vectors} through multiple \textit{hidden layers}, with the latter being functions applying successive
	affine and nonlinear transformations to input vectors.
	In this paper, the type of neural network considered to represent global hedging policies is the \textit{long short-term memory} (LSTM, \cite{hochreiter1997long}). LSTMs belong to the class of \textit{recurrent neural networks} (RNNs, \cite{rumelhart1986learning}), which have self-connections in hidden layers: the output of the time-$t_n$ hidden layer is a function of both the time-$t_n$ feature vector as well as the output of the time-$t_{n-1}$ hidden layer. The periodic computation of long short-term memory neural networks is done with so-called \textit{LSTM cells}, which are similar to but more complex than the typical hidden layer of RNNs. 
	LSTMs have recently been applied with success to approximate global hedging policies in several studies: \cite{buehler2019deep_2}, \cite{cao2020discrete} and \cite{carbonneau2020deep}. 
	Additional remarks are made in subsequent sections to motivate this choice of neural networks for the specific setup of this paper. 
	
	\subsection{Neural networks representing trading policies}
	\label{subsec:neural_net_hedging}
	The following formally defines the architecture of long-short term memory neural networks. For convenience, a very similar notation for neural networks as the one of \cite{carbonneau2020deep} is used. Note that the time steps of the feature and output vectors coincide with the set of financial market trading dates $\mathcal{T}$. For additional general information about LSTMs, the reader is referred to Chapter $10.10$ of \cite{goodfellow2016deep} and the many references therein. 

	\begin{Def}{(LSTM)}
		\label{def:LSTM}
		For $H, d_0, \ldots, d_{H+1} \in \mathbb{N}$, let $F_{\theta}:\mathbb{R}^{N \times d_0} \rightarrow \mathbb{R}^{N \times d_{H+1}}$ be an LSTM which maps the sequence of feature vectors $\{X_{n}\}_{n=0}^{N-1}$ to output vectors $\{Y_{n}\}_{n=0}^{N-1}$ where $X_{n} \in \mathbb{R}^{d_0}$ and $Y_{n} \in \mathbb{R}^{d_{H+1}}$ for $n=0,\ldots,N-1$. 
		The computation of $Y_n$, the subset of outputs of $F_{\theta}$ associated with time $t_n$, is achieved through $H$ LSTM cells, 
		each of which
		outputs a vector of $d_{j}$ neurons denoted as $h_{n}^{(j)} \in \mathbb{R}^{d_{j} \times 1}$ for $j=1,\ldots,H$. More precisely, the computation applied by the $j^{\text{th}}$ LSTM cell for the time-$t_{n}$ output is as
		follows:\footnote{
			At time $0$ (i.e. $n=0$), the computation of the LSTM cells is the same as in \eqref{eq:ref_LSTM_cell} with $h_{-1}^{(j)}$ and $c_{-1}^{(j)}$ defined as vectors of zeros of dimension $d_{j}$ for $j=1,\ldots,H$.
		} 
		\begin{align}
			i_{n}^{(j)} &= \text{sigm}(U_i^{(j)} h_n^{(j-1)} + W_i^{(j)} h_{n-1}^{(j)} + b_i^{(j)}), \nonumber
			%
			\\ f_{n}^{(j)} &= \text{sigm}(U_f^{(j)} h_n^{(j-1)} + W_f^{(j)} h_{n-1}^{(j)} + b_f^{(j)}), \nonumber
			\\ o_{n}^{(j)} &= \text{sigm}(U_o^{(j)} h_n^{(j-1)} + W_o^{(j)} h_{n-1}^{(j)} + b_o^{(j)}), \nonumber
			\\ c_{n}^{(j)} &= f_{n}^{(j)} \odot c_{n-1}^{(j)} + i_{n}^{(j)} \odot \text{tanh}(U_c^{(j)} h_n^{(j-1)} + W_c^{(j)} h_{n-1}^{(j)} + b_c^{(j)}), \nonumber
			\\ h_{n}^{(j)} &= o_{n}^{(j)} \odot \text{tanh}(c_{n}^{(j)}), \label{eq:ref_LSTM_cell}
		\end{align}
		where $\odot$ denotes the Hadamard product (the element-wise product), $\text{sigm}(\cdot)$ and $\text{tanh}(\cdot)$ are the sigmoid and hyperbolic tangent functions applied element-wise to each scalar given as input\footnote{
			For $X:=[X_1,\ldots,X_K]$, $\text{sigm}(X) := \left[\frac{1}{1 + e^{-X_1}}, \ldots, \frac{1}{1 + e^{-X_K}}\right]$ and $\text{tanh}(X) := \left[\frac{e^{X_1}-e^{-X_1}}{e^{X_1}+e^{-X_1}}, \ldots, \frac{e^{X_K}-e^{-X_K}}{e^{X_K}+e^{-X_K}}\right]$.
		} and
		\begin{itemize}
			\item $U_i^{(j)}, U_f^{(j)}, U_o^{(j)}, U_c^{(j)} \in \mathbb{R}^{d_j  \times d_{j-1}}$, $W_i^{(j)}, W_f^{(j)}, W_o^{(j)}, W_c^{(j)} \in \mathbb{R}^{d_j  \times d_{j}}$ and $b_i^{(j)}, b_f^{(j)}, b_o^{(j)}, b_c^{(j)} \in \mathbb{R}^{d_j \times 1}$ for $j=1,\ldots,H$. 
		\end{itemize}
		At each time-step, the input of the first LSTM cell is the feature vector $($i.e. $h_{n}^{(0)}:=X_{n}$$)$ and the final output is an affine transformation of the output of the last LSTM cell:
		\begin{align}
			Y_{n} = W_{y}h_{n}^{(H)} + b_{y}, \quad n = 0,\ldots,N-1, \label{eq:ref_output_LSTM}
		\end{align}
		where $W_y \in \mathbb{R}^{d_{H+1} \times d_{H}}$ and $b_y \in \mathbb{R}^{d_{H+1} \times 1}$. Lastly, the set of trainable parameters denoted as $\theta$ consists of all weight matrices and bias vectors:
		\begin{align}
			\theta := \left\{\{U_i^{(j)}, U_f^{(j)}, U_o^{(j)}, U_c^{(j)}, W_i^{(j)}, W_f^{(j)}, W_o^{(j)}, W_c^{(j)}, b_i^{(j)}, b_f^{(j)}, b_o^{(j)}, b_c^{(j)}\}_{j=1}^{H}, W_y, b_y\right\}. \label{eq:ref_trainable_params}
		\end{align}
	\end{Def}
	
	In this study, the computation of hedging positions is done through the mapping of a sequence of relevant financial market observations into the periodic number of shares held in each hedging instrument with an LSTM. One of the main objectives of this paper is to analyze the impact of including vanilla options as hedging instruments on the ERP framework.
	For the numerical experiments conducted in the subsequent \cref{section:numerical_results}, the hedging instruments consist of either only the underlying asset (without options) or exclusively options (without the underlying asset). The case of using both the stock and options is not considered since the options can always replicate positions in the underlying asset with calls and puts by relying on the put-call parity.
	%
	In what follows, let $\{X_n\}_{n=0}^{N-1}$ and $\{Y_n\}_{n=0}^{N-1}$ be respectively the sequence of feature vectors and output vectors of an LSTM as in \cref{def:LSTM}. When hedging only with the underlying, the time-$t_n$ feature vector considered is\footnote{
		The use of $\log(S_{n}^{(0,b)}/K)$ instead of $S_{n}^{(0,b)}$ in feature vectors was found to improve the learning speed of the neural networks (i.e. time taken to find a good set of trainable parameters). Note that log transformation for risky asset prices was also considered in \cite{carbonneau2020deep}, \cite{buehler2019deep} and \cite{buehler2019deep_2}.
		%
	}
	\begin{align}
		X_{n} = [\log(S_{n}^{(0,b)}/K), V_{n}^{\delta}, \varphi_n], \quad n =0,\ldots,N-1, \label{eq:ref_feature_vect_stock}
	\end{align}
	where $K$ is the strike price of $\Phi$ and $\{\varphi_n\}_{n=0}^{N-1}$ is a sequence of additional relevant state variables associated with the dynamics of asset prices. For instance, if the underlying log-returns are modeled with a GARCH process, it is well-known that the bivariate process of the underlying price and the GARCH volatility has the Markov property under $\mathbb{P}$ with respect to the market filtration $\mathbb{F}$. The time-$t_n$ volatility of the GARCH process is thus added to the feature vectors through $\varphi_n$. 
	Furthermore, in that same case where the underlying stock is considered as the only hedging instrument, the output vectors of the LSTM consist of the number of underlying asset shares to be held in the portfolio for all time steps, i.e. $Y_{n} = \delta_{n+1}^{(0)}$ for $n=0,\ldots,N-1$. 
	
	Conversely, when hedging is performed with options as hedging instruments, the \textit{implied volatilities} (IVs) of such options denoted as $\{IV_{n}\}_{n=0}^{N-1}$ are added to feature vectors with $IV_n$ encompassing every implied volatilities needed to price the $D$ options used for hedging:\footnote{
		Note that the bijection relation between implied volatilities and option prices entails that either values could theoretically be used in feature vectors as one is simply a nonlinear transformation of the other. 
	}
	\begin{align}
		X_{n} = [\log(S_{n}^{(0,b)}/K), V_{n}^{\delta}, \varphi_n, IV_{n}], \quad n =0,\ldots,N-1. \label{eq:ref_feature_vect_opts}
	\end{align}
	In that case, the output vectors are the number of option contracts held in the portfolio for the various time steps: $Y_{n}=[\delta_{n+1}^{(1)}, \ldots, \delta_{n+1}^{(D)}]$ for $n=0,\ldots,N-1$. Recall that when options are used as hedging instruments, $\delta_{n+1}^{(0)}=0$ for $n=0,\ldots,N-1$.
	\begin{remark}
		\label{remark:adding_port_value}
		Although the portfolio value $V_{n}^{\delta}$ 
		is in theory
		a redundant feature in the context of LSTMs since it can be retrieved as a function of previous times inputs and outputs of the neural network (see \eqref{eq:ref_update_rule_port_value}), incorporating it to feature vectors was found to significantly improve upon the hedging effectiveness of the LSTMs in the numerical experiments conducted in \cref{section:numerical_results}. 
	\end{remark}
	
	\subsection{Equal risk pricing with neural networks}
	\label{subsec:deep_ERP}
	To numerically solve the underlying global hedging problems of the ERP framework, \cite{carbonneau2020equal} propose to use two distinct neural networks denoted as $F_{\theta}^{(\mathcal{L})}$ and $F_{\theta}^{(\mathcal{S})}$ to approximate the global trading policies of respectively the long and short positions in $\Phi$. This is the approach considered in the current paper. As illustrated below, the procedure consists in solving the alternative problems of optimizing the neural networks trainable parameters so as to minimize the corresponding hedging shortfall:
	\begin{align}
		\epsilon^{(\mathcal{L})}(V_0) &\approx \underset{\theta \in \mathbb{R}^{q}}{\min} \, \rho \left(-\Phi(S_{N}^{(0,b)}) -B_{N}(V_0 + G_{N}^{\delta^{(\mathcal{L},\theta)}})\right), \label{eq:risk_long_NNET}
		\\
		\epsilon^{(\mathcal{S})}(V_0) &\approx \underset{\theta \in \mathbb{R}^{q}}{\min} \, \rho \left(\Phi(S_{N}^{(0,b)}) - B_{N}(V_0 + G_{N}^{\delta^{(\mathcal{S},\theta)}})\right), \label{eq:risk_short_NNET}
	\end{align}	
	where $\delta^{(\mathcal{L},\theta)}$ and $\delta^{(\mathcal{S},\theta)}$ are to be understood respectively as the output sequences of $F_{\theta}^{(\mathcal{L})}$ and $F_{\theta}^{(\mathcal{S})}$, and $q \in \mathbb{N}$ is the total number of trainable parameters of $F_{\theta}^{(\mathcal{L})}$ and $F_{\theta}^{(\mathcal{S})}$.
	The approximated measured risk exposures obtained through \eqref{eq:risk_long_NNET} and \eqref{eq:risk_short_NNET} are subsequently used to compute equal risk prices and $\epsilon$-completeness measures with \eqref{eq:ref_equal_risk_prices} and \eqref{eq:ref_epsilon_def}.
	%
	One implicit assumption associated with \eqref{eq:risk_long_NNET} and \eqref{eq:risk_short_NNET} is that the architecture of all neural networks in terms of the number of LSTM cells and neurons per cell is always fixed; the hyperparameter tuning step of the optimization problem is not considered in this paper. \cref{subsec:SGD} that follows presents the procedure considered in this study to optimize the trainable parameters of the LSTMs. 
	
	\begin{remark}
		\label{remark:why_LSTM}
		\cite{carbonneau2020equal} show that when relying on feedforward neural networks (FFNNs\footnote{FFNNs are another class of neural networks  which map input vectors into output vectors, in contrast to LSTMs which map input vector sequences to output vector sequences.}) instead of LSTMs, the alternative problems \eqref{eq:risk_long_NNET}-\eqref{eq:risk_short_NNET} allow for arbitrarily precise approximations of the measured risk exposures \eqref{eq:risk_long}-\eqref{eq:risk_short} due to results from \cite{buehler2019deep}. Despite this theoretical ability of FFNNs to approximate arbitrarily well global hedging policies in such context, the authors of the current paper found that LSTMs are able to learn significantly better trading policies than FFNNs in the numerical experiments carried out in \cref{section:numerical_results}, which motivates their use over FFNNs. The theoretical justifications for the outperformance of LSTMs over FFNNs in the financial market settings of this paper are out-of-scope and are left-out as interesting potential future research. 
	\end{remark}

	\subsection{Training neural networks}
	\label{subsec:SGD}
	The numerical scheme to optimize the trainable parameters of neural networks as entailed by the global hedging optimization problems \eqref{eq:risk_long_NNET}-\eqref{eq:risk_short_NNET} is now described. The procedure first proposed in \cite{buehler2019deep} uses minibatch stochastic gradient descent (SGD) to approximate the gradient of the cost function with Monte Carlo sampling. For convenience, the notation used for the optimization procedure is similar to the one from \cite{carbonneau2020equal}. Without loss of generality, the numerical procedure is only presented for the short measured risk exposure; the corresponding procedure for the long position is simply obtained through modifying the objective function \eqref{eq:ref_cost_func_short} that follows. Let $J:\mathbb{R}^{q} \rightarrow \mathbb{R}$ be the cost function to be minimized for the short position in $\Phi$, where $\theta$ is the set of trainable parameters of $F_{\theta}^{(\mathcal{S})}$:\footnote{
	Minimizing $J$ with respect to $\theta$ corresponds to the alternative problem \eqref{eq:risk_short_NNET} with zero initial capital. Recall that $\epsilon^{(\mathcal{L})}(0)$ and $\epsilon^{(\mathcal{S})}(0)$ are required for the computation of $C_0^{\star}$ and $\epsilon^{\star}$. Consequently, hedging portfolio values used in LSTM feature vectors are equal to hedging gains, i.e. $V_n^{\delta} = B_nG_n^{\delta}$.
	}
	\begin{align}
		J(\theta):= \rho \left(\Phi(S_{N}^{(0,b)}) - B_{N}G_{N}^{\delta^{(\mathcal{S},\theta)}}\right), \quad \theta \in \mathbb{R}^{q}. \label{eq:ref_cost_func_short}
	\end{align}
	A typical stochastic gradient descent procedure entails adapting the trainable parameters iteratively and incrementally in the opposite direction of the cost function gradient with respect to $\theta$:
	\begin{align}
		\theta_{j+1} &= \theta_{j} - \eta_j \nabla_{\theta}J(\theta_{j}), \label{eq:ref_SGD_iterative_param}
	\end{align}
	where $\theta_0$ is the initial values for the trainable parameters,
	$\eta_j$ is a small deterministic positive real value commonly called the \textit{learning rate} and $\nabla_{\theta}$ denotes the gradient operator. In the current study, the \textit{Glorot uniform initialization} of \cite{glorot2010understanding} is always used to select initial parameters in $\theta_0$.
	Since closed-form solutions for the gradient of the cost function with respect to the trainable parameters are 
 	unavailable
	in the general market setting considered in this work, the approach relies instead on Monte Carlo sampling 
 	to provide an estimate. Thus, let $\mathbb{B}_j:=\{\pi_{i,j}\}_{i=1}^{N_{\text{batch}}}$ be a minibatch of simulated hedging errors of size $N_{\text{batch}} \in \mathbb{N}$ where $\pi_{i,j}$ is the $i^{\text{th}}$ simulated hedging error when $\theta = \theta_j$:
	\begin{align}
		\pi_{i,j}:= \Phi(S_{N,i}^{(0,b)}) - B_{N}G_{N,i}^{\delta^{(\mathcal{S},\theta_j)}}, \label{eq:ref_simulated_hedging_error}
	\end{align}  
	where $S_{N,i}^{(0,b)}$ and $G_{N,i}^{\delta^{(\mathcal{S},\theta_j)}}$ are the $i^{th}$ random realization among the minibatch of the terminal underlying asset price and discounted hedging portfolio gains, respectively.
	Furthermore, denote $\hat{\rho}:\mathbb{R}^{N_{\text{batch}}} \rightarrow \mathbb{R}$ as the empirical estimator of $\rho(\Phi(S_{N}^{(0,b)}) - B_{N}G_{N}^{\delta^{(\mathcal{S},\theta)}})$ evaluated with minibatches of hedging errors. 
	Minibatch SGD consists in approximating the gradient of the cost function $\nabla_{\theta}J(\theta_{j})$ with $\nabla_{\theta}\hat{\rho}(\mathbb{B}_j)$ in the update rule for trainable parameters:
	\begin{align}
		\theta_{j+1} &= \theta_{j} - \eta_j \nabla_{\theta}\hat{\rho}(\mathbb{B}_j). \label{eq:ref_SGD_iterative_param_approx}
	\end{align}
	
	For the numerical experiments conducted in \cref{section:numerical_results}, the convex risk measure considered is the Conditional Value-at-Risk (CVaR, \cite{rockafellar2002conditional}). For an absolutely continuous integrable random variable\footnote{
		In \cref{section:numerical_results}, the only dynamics considered for the risky assets produce integrable and absolutely continuous hedging errors. 
	}, the CVaR has the representation
	\begin{align}
		\text{CVaR}_{\alpha}(X):=\E[X|X \geq \text{VaR}_{\alpha}(X)], \quad \alpha \in (0,1), \label{eq:ref_CVaR_def}
	\end{align}
	where $\text{VaR}_{\alpha}(X) := \min \left\{x : \mathbb{P}(X \leq x) \geq \alpha \right\}$ is the Value-at-Risk (VaR) with confidence level $\alpha$ of the liability $X$. 
	Let $\{\pi_{[i],j}\}_{i=1}^{N_{\text{batch}}}$ be the order statistics (i.e. values sorted by increasing order) of $\mathbb{B}_{j}$. For $\tilde{N}:= \ceil*{\alpha N_{\text{batch}}}$ where $\ceil{x}$ is the ceiling function (i.e. the smallest integer greater or equal to $x$), the empirical estimator of the CVaR used in this study is from the work of \cite{hong2014monte} 
	and has the representation
	$$\reallywidehat{\text{VaR}}_{\alpha}(\mathbb{B}_{j}) := \pi_{[\tilde{N}], j},$$
	$$\reallywidehat{\text{CVaR}}_{\alpha}(\mathbb{B}_{j}):= \reallywidehat{\text{VaR}}_{\alpha}(\mathbb{B}_{j}) + \frac{1}{(1-\alpha)N_{\text{batch}}}\sum_{i=1}^{N_{\text{batch}}}\max(\pi_{i,j}-\reallywidehat{\text{VaR}}_{\alpha}(\mathbb{B}_{j}),0).$$
	The gradient of the empirical estimator of the Conditional Value-at-Risk with respect to the trainable parameters (i.e. $\nabla_{\theta}\reallywidehat{\text{CVaR}}_{\alpha}(\mathbb{B}_{j})$) required for the update rule \eqref{eq:ref_SGD_iterative_param_approx} can be computed exactly without discretization or other numerical approximations. Such computations can be implemented with modern deep learning libraries such as Tensorflow \citep{abadi2016tensorflow}. 
	Furthermore, algorithms which dynamically adapt the learning rate $\eta_j$
	in \eqref{eq:ref_SGD_iterative_param_approx} such as \textit{Adam} \citep{kingma2014adam} have been shown to improve upon the effectiveness of SGD procedures for neural networks. For all numerical experiments conducted in \cref{section:numerical_results}, an implementation of Tensorflow with the Adam algorithm is used to optimize neural networks; the reader is referred to the online Github repository for samples of codes in Python.\footnote{
		\href{https://github.com/alexandrecarbonneau}{github.com/alexandrecarbonneau}.
	} Also, \cref{appendix:pseudo_code} presents a pseudo-code of the training procedure for $F_{\theta}^{\mathcal{(S)}}$.
	

	\section{Numerical experiments}
	\label{section:numerical_results}
	This section performs various numerical experimentations of the ERP approach for derivatives valuation. The main goal is to study the impact of including options as hedging instruments on equal risk prices and on the level of market incompleteness. A special case assessed throughout this section is 
	the trading of short-term vanilla options for the pricing and hedging of longer-term derivatives. The conduction of these experiments heavily relies on the neural network scheme described in \cref{section:neural_network} to solve the underlying global hedging problems of the ERP framework. Such exhaustive numerical study would have been hardly accessible with traditional methods (e.g. conventional dynamic programming algorithms) due to the high-dimensional continuous state and action spaces of the hedging problem stemming from the use of multiple short-term options as hedging instruments and from the asset price dynamics considered. As a result, the use of neural networks enables us to provide novel qualitative insights into the ERP framework.

	The analysis begins in \cref{subsec:sensitivity_MJD} and \cref{subsec:sensitivity_GARCH} with the assessment of the sensitivity of equal risk prices and residual hedging risk to the presence of two salient equity stylized features: jump and volatility risks. The impact of the choice of convex risk measure on the ERP framework when trading exclusively options is examined in \cref{subsec:sensitivity_risk_aversion}. Lastly, \cref{subsec:benchmark_QDH_ERP} presents the benchmarking of equal risk prices to derivative premiums obtained with variance-optimal hedging. The specific financial market setup and asset dynamics models considered for all numerical experiments are described in \cref{subsec:financial_market} that follows.

	\subsection{Market setup and asset dynamics models}
	\label{subsec:financial_market}
	For the rest of the paper, the derivative to price is a European vanilla put option of payoff function $\Phi(S_N^{(0,b)}) = \max(K-S_N^{(0,b)},0)$ with $K = 90, 100$ and $110$ corresponding respectively to an out-of-the-money (OTM), an at-the-money (ATM) and an in-the-money (ITM) option. The maturity of the derivative is set to $1$ year (i.e. $T=1$) with $252$ days. The annualized continuously compounded risk-free rate is $r=0.03$. In addition to the risk-free asset, the hedging instruments consist of either only the underlying stock, or exclusively shorter-term ATM European calls and puts. When hedging is performed with the underlying stock, daily and monthly rebalancing are considered, corresponding to respectively $N=252$ and $N=12$ trading periods per year. When hedging with options, all options are assumed to have a single-period time-to-maturity, i.e. they are traded once and held until expiration. We consider either 1-month or 3-months maturities ATM calls and puts as hedging instruments, which respectively entails $N=12$ or $N=4$. Less frequent rebalancing when hedging with options rather than only with the underlying stock is consistent with market practices; such hedging instruments are commonly embedded in semi-static type of trading strategies, see for instance \cite{carr2014static}. Lastly, note that daily variations for the underlying log-returns and implied volatilities are always considered throughout the rest of the paper, even with non-daily rebalancing periods (i.e. when hedging with the underlying stock on a monthly basis or with $1$-month and $3$-months maturities options) by aggregating daily variations over the rebalancing period.

	\subsubsection{Asset price dynamics}
	\label{subsec:dynamics_underlying}
	The asset price dynamics models considered in stochastic simulations are now introduced. To characterize jump risk, the Merton jump-diffusion model (MJD, \cite{Merton1976}) is considered. Furthermore, the impact of volatility risk is assessed with the GJR-GARCH model of \cite{glosten1993relation}. Several sets of parameters are tested for each model to conduct a sensitivity analysis and highlight the impact of various model features on both equal risk prices and residual hedging risk. 
	%

	Denote $y_n:=\log (S_n^{(0,b)}/S_{n-1}^{(0,b)})$ as the periodic underlying stock log-return between the trading periods $t_{n-1}$ and $t_{n}$. Since our modeling framework assumes daily variations for asset prices and possibly non-daily rebalancing, let $\{\tilde{y}_{j,n}\}_{j=1}^{M}$ be the $M$ daily stock log-returns in the time interval $[t_{n-1}, t_{n}]$ such that\footnote{
		For completeness, let $\{\tilde{S}^{(0,b)}_{j,n}\}_{j=0,n=1}^{M,N}$ be the daily underlying stock prices where $\{\tilde{S}^{(0,b)}_{j,n}\}_{j=0}^{M}$ corresponds to the $M+1$ daily prices during the period $[t_{n-1}, t_n]$. Also, let $\mathbb{G}:=\{\mathcal{G}_{j,n}\}_{j=0, n=1}^{M,N}$ be a filtration satisfying the usual conditions with $\mathcal{G}_{j,n}$ containing all information available to market participants at the $j^{th}$ day of the time period $[t_{n-1},t_n]$. The filtration used to optimize trading strategies $\mathbb{F}$ with time steps $t_0, t_1,\ldots,t_N$ is a subset of $\mathbb{G}$ by construction. However, since the risky asset dynamics considered in this paper have the Markov property, optimizing trading strategies with the filtration $\mathbb{F}$ or $\mathbb{G}$ results in the same trading policy.
	}
	\begin{align}
	y_n = \sum_{j=1}^{M}\tilde{y}_{j,n}, \quad n = 1,\ldots,N, \quad N \times M = 252, \label{eq:ref_agg_return}
	\end{align}
	where $N$ corresponds to the number of trading dates to hedge the $1$ year maturity derivative $\Phi$ and $M$ to the number of days between two trading dates. Thus, daily stock hedges corresponds to the case of $N=252$ and $M=1$, monthly stock and $1$-month option hedges to $N=12$ and $M=21$, and $3$-months option hedges to $N=4$ and $M=63$. 

	The asset price dynamics are now formally defined for the daily log-returns. For the rest of the section, let $\{\epsilon_{j,n}\}_{j=1, n=1}^{M,N}$ be a sequence of independent standardized Gaussian random variables where the subsequence $\{\epsilon_{j,n}\}_{j=1}^{M}$ will be used to model the $M$ daily innovations of log-returns in the time interval $[t_{n-1}, t_n]$. 

	\subsubsection{Discrete-time Merton-Jump diffusion model \citep{Merton1976}}
	\label{subsubsec:MJD_model}
	The Merton-jump diffusion dynamics expands upon the ideal market conditions of the Black-Scholes model by incorporating random Gaussian jumps along stock paths. Let $\{N_{j,n}\}_{j=0,n=1}^{M,N}$ be a discrete-time sampling from a Poisson process of intensity parameter $\lambda > 0$ where the subsequence $\{N_{j,n}\}_{j=0}^{M}$ corresponds to the $M+1$ daily values of the Poisson process occurring during the time interval $[t_{n-1}, t_n]$. $N_{0,1}:=0$ is the initial value of the process and $N_{0,n+1} := N_{M,n}$ for $n=1,\ldots,N-1$. Furthermore, denote $\{\xi_k\}_{k=1}^{\infty}$ as a sequence of random Gaussian variables corresponding to the jumps of mean $\mu_{J}$ and variance $\sigma_{J}^{2}$. $\{N_{j,n}\}_{j=0,n=1}^{M,N}$, $\{\xi_k\}_{k=1}^{\infty}$ and $\{\epsilon_{j,n}\}_{n=1,j=1}^{N,M}$ are independent. For $n=1,\ldots,N$ and $j=1,\ldots, M$, the daily log-return dynamics can be specified as\footnote{
	This paper adopts the convention that if $N_{j,n}=N_{j-1,n}$, i.e. that no jumps occurred on that day, then:
	$$\sum_{k=N_{j-1,n} + 1}^{N_{j,n}}\xi_{k} = 0.$$ 
}
\begin{align}
\tilde{y}_{j,n} = \frac{1}{252}\left(\nu - \lambda \left(e^{\mu_J + \sigma_J^{2}/2}-1\right) -  \frac{\sigma^{2}}{2}\right) + \sigma \sqrt{\frac{1}{252}}\epsilon_{j,n} + \sum_{k=N_{j-1,n} + 1}^{N_{j,n}}\xi_{k}, \label{eq:ref_MJD}
\end{align}
where $\{\nu, \mu_J, \sigma_J, \lambda, \sigma\}$ are the model parameters with $\{\nu, \lambda, \sigma\}$ being on a yearly scale, $\nu \in \mathbb{R}$ and $\sigma > 0$. Furthermore, since $\{S_n^{(0,b)}\}_{n=0}^{N}$ has the Markov property with respect to the filtration $\mathbb{F}$ generated by the trading dates observations, no additional state associated to the risky asset dynamics is required to be added to the feature vectors of neural networks
(i.e. $\varphi_n = 0$ for all time steps $n$ in \eqref{eq:ref_feature_vect_stock} and \eqref{eq:ref_feature_vect_opts}).

\subsubsection{GJR-GARCH(1,1) model \citep{glosten1993relation}}
\label{subsubsec:GARCH_model}
GARCH processes also expand upon the Black-Scholes ideal framework by exhibiting well-known empirical features of risky assets such as time-varying volatility, volatility clustering and the leverage effect (i.e. negative correlation between underlying returns and its volatility). Daily log-returns modeled with a GJR-GARCH(1,1) dynamics have the representation
\begin{align}
\tilde{y}_{j,n} &= \mu + \tilde{\sigma}_{j,n} \epsilon_{j,n}, \nonumber
\\
\tilde{\sigma}_{j+1, n}^{2} &= \omega + \upsilon \tilde{\sigma}_{j,n}^{2}(|\epsilon_{j,n}| - \gamma \epsilon_{j,n})^{2} + \beta \tilde{\sigma}_{j,n}^{2}, \label{eq:ref_GARCH}	
\end{align}
where $\{\tilde{\sigma}_{j,n}^2\}_{j=1,n=1}^{M+1,N}$ are the daily conditional variances of log-returns. More precisely, $\{\tilde{\sigma}_{j,n}^2\}_{j=1}^{M+1}$ are the $M+1$ daily conditional variances in the time interval $[t_{n-1}, t_n]$. Also, $\tilde{\sigma}_{1,n+1}^2 := \tilde{\sigma}_{M+1,n}^2$ for $n=1,\ldots,N-1.$ Model parameters consist of $\{\mu, \omega, \upsilon, \gamma, \beta\}$ with $\{\omega, \upsilon, \beta\}$ being positive real values and $\gamma, \mu \in \mathbb{R}$. Note that if the starting value of the GARCH process $\tilde{\sigma}_{1,1}^{2}$ is deterministic, then $\{\tilde{\sigma}_{j,n}^2\}_{j=1,n=1}^{M+1,N}$ can be computed recursively with the observed daily log-returns. In this paper, $\tilde{\sigma}_{1,1}^{2}$ is set as the stationary variance: $\tilde{\sigma}_{1,1}^{2} := \frac{\omega}{1 - \upsilon(1+\gamma^{2})-\beta}$. Also, contrarily to the MJD model, the GJR-GARCH(1,1) requires adding at each trading time $t_n$ the current stochastic volatility value to the feature vectors of the neural networks, i.e. $\varphi_n = \tilde{\sigma}_{1,n+1}$ for $n=0,\ldots,N-1$.  

\subsubsection{Implied volatility dynamics}
\label{subsec:dynamics_IV}	
This work proposes to model the daily variations of the logarithm of ATM implied volatilities with 1-month and 3-months maturities as a discrete-time version of the Ornstein-Uhlenbeck (OU) process.\footnote{
	It is important to note that implied volatilities are used strictly for pricing options used as hedging instruments. They are not used to price the derivative $\Phi$. 
} 
The choice of an OU type of dynamics for IVs is motivated by the work of \cite{cont2002dynamics} which shows that for S\&P 500 index options,
the first principal component
of the daily variations of the logarithm of the IV surface accounts for the majority of its variance and can be interpreted as a level effect. Also, this first principal component can be well represented by a low-order autoregressive (AR) model. 
The OU dynamics considered in this study therefore has the representation of an AR model of order $1$.
%

The dynamics for the daily evolution of IVs is now formally defined. For convenience, this paper assumes that 1-month and 3-months IVs are the same.\footnote{
It is worth highlighting that since trading strategies allow for the use of either $1$-month or $3$-months maturities ATM calls and puts, but not both maturities within the same strategy, $1$-month and $3$-months IVs are never used at the same time.
} 
Using a similar notation as for daily log-returns, let $\{\widetilde{IV}_{j,n}\}_{j=0,n=1}^{M,N}$ be the daily ATM IV process for both 1-month and 3-months maturities where $\{\widetilde{IV}_{j,n}\}_{j=0}^{M}$ are the $M+1$ daily observations during the time interval $[t_{n-1}, t_n]$ with $\widetilde{IV}_{0,n+1}:=\widetilde{IV}_{M,n}$ for $n=1,\ldots,N-1$. 
Furthermore, let $\{Z_{j,n}\}_{j=1,n=1}^{M,N}$ be an additional sequence of independent standardized Gaussian random variables characterizing shocks in the IV dynamics.
In order to incorporate the stylized feature of strong negative correlation between implied volatilities and asset returns (\cite{cont2002dynamics}), the modeling framework assumes that the daily innovations of log-returns and IVs are correlated with parameter $\varrho := corr(\epsilon_{j,n},Z_{j,n})$ set at $-0.6$ for all time steps.
The dynamics for the evolution of the logarithm of IVs, which is referred from now on as the log-AR(1) model, has the following representation for $n=1,\ldots,N$ and $j=0,\ldots,M-1$:
\begin{align}
\log \widetilde{IV}_{j+1,n} &= \log \widetilde{IV}_{j,n} + \kappa(\vartheta  - \log \widetilde{IV}_{j,n}) + \sigma_{IV} Z_{j+1,n}, \label{eq:ref_IV_model}
\end{align}
where $\{\kappa, \vartheta , \sigma_{IV}\}$ are the model parameters with $\kappa, \vartheta \in \mathbb{R}$ and $\sigma_{IV} > 0$. The initial value of the process is set as $\log \widetilde{IV}_{0,1} = \vartheta$. Also, recall that when trading options, their corresponding implied volatilities at each trading date are added to the feature vectors of neural networks, i.e. $IV_{n-1} = \widetilde{IV}_{0,n}$ in \eqref{eq:ref_feature_vect_opts} for $n=1,\ldots,N$.

The pricing of calls and puts used as hedging instruments is done with the well-known Black-Scholes formula hereby stated with the annual volatility term set at the implied volatility value. For the underlying price $S$, implied volatility $IV$, strike price $K$ and time-to-maturity $\Delta T$, the Black-Scholes pricing formulas for calls and puts are respectively
\begin{align}
C(S, IV, \Delta T, K)&:=S \mathcal{N}(d_1) - e^{-r\Delta T}K \mathcal{N}(d_2), \label{eq:ref_BSM_optprice_call}
\\ P(S, IV, \Delta T, K)&:=e^{-r\Delta T} K \mathcal{N}(-d_2) - S \mathcal{N}(-d_1), \label{eq:ref_BSM_optprice_put}
\end{align}
where $\mathcal{N}(\cdot)$ denotes the standard normal cumulative distribution function and
$$d_1:=\frac{\log(\frac{S}{K}) + (r+\frac{IV^2}{2})\Delta T}{IV\sqrt{\Delta T}}, \quad d_2:=d_1 - IV\sqrt{\Delta T}.$$

\subsubsection{Hyperparameters}
\label{subsubsec:hyperparameters}
The set of hyperparameters for the LSTMs are two LSTM cells (i.e. $H=2$) and $24$ neurons per cell (i.e. $d_j = 24$ for $j=1,2$). A training set of $400,\!000$ paths is used to optimize the trainable parameters with a total of $50$ epochs and a minibatch size of $1,\!000$ sampled exclusively from the training set.\footnote{
	One epoch consists of a complete iteration of SGD on the training set. For a training set of $400,\!000$ paths and a minibatch of size $1,\!000$, a total of $400$ updates of the trainable parameters as in \eqref{eq:ref_SGD_iterative_param_approx} is performed within an epoch.
} The deep learning library Tensorflow \citep{abadi2016tensorflow} is used to implement the stochastic gradient descent procedure with the Adam optimizer of \cite{kingma2014adam} with a learning rate hyperparameter value of $0.01/6$. All numerical results presented throughout this section are computed based on a test set (i.e. out-of-sample dataset) of $100,\!000$ paths. Lastly, unless specified otherwise, the convex risk measure chosen for all experiments is the CVaR with confidence level $\alpha = 0.95$. Sensitivity analyses of equal risk prices and residual hedging risk with respect to the confidence level parameter of the CVaR measure are performed in \cref{subsec:sensitivity_risk_aversion} and \cref{subsec:benchmark_QDH_ERP}.


\subsection{Sensitivity of equal risk pricing to jump risk}
\label{subsec:sensitivity_MJD}

This section examines the sensitivity of the ERP solution to equity jump risk.
The analysis is carried out by considering three different sets of parameters for the MJD dynamics which induce different levels of jump frequency and severity. While maintaining empirical plausibility, this is done by modifying the intensity parameter $\lambda$ controlling the expected frequency of jumps as well as parameters $\mu_J$ and $\sigma_J$ controlling the severity component of jumps. In order to better isolate the impact of different stylized features of jump risk on the ERP framework, the diffusion parameter\footnote{
	The parameter $\sigma$ in \eqref{eq:ref_MJD} corresponds to the diffusion parameter of the MJD dynamics.
} is fixed for all three sets of parameters. Also, the parameters $\{\lambda, \mu_J, \sigma_J, \nu\}$ are chosen such that the yearly expected value and standard deviation of log-returns are respectively $10\%$ and $15\%$ for all three cases. To facilitate the analysis, the three sets of parameters are referred to as scenario $1$, scenario $2$ and scenario $3$ for jump risk. Model parameter values for the three scenarios are presented in \cref{table:all_MJD}.
\begin{table}[ht]
	\caption {Parameters of the Merton jump-diffusion model for the three scenarios.} \label{table:all_MJD}
	\begin{adjustwidth}{-1in}{-1in} 
		\centering  
		\begin{tabular}{lccccc}
			\hline
			$ $ & $\nu$ & $\sigma$ & $\lambda$ & $\mu_J$ & $\sigma_J$
			\\
			\hline\noalign{\medskip}
			$\text{Scenario 1}$ & $0.1112$  &  $0.1323$ & $1$ & $-0.05$ & $0.05$
			\\
			%
			$\text{Scenario 2}$ & $0.1111$  &  $0.1323$ & $0.25$ & $-0.10$ & $0.10$
			\\
			%
			$\text{Scenario 3}$ & $0.1110$  &  $0.1323$ & $0.08$ & $-0.20$ & $0.15$
			\\    
			\noalign{\medskip}\hline
		\end{tabular}%
	\end{adjustwidth}
	\centering{Notes: $\nu$, $\sigma$ and $\lambda$ are on an annual basis.}
\end{table}
Scenario $1$ represents relatively smaller but more frequent jumps with on average one jump per year of mean $-5\%$ and standard deviation $5\%$. Scenario $2$ entails more severe, but less frequent jumps with on average one jump every four years of mean $-10\%$ and standard deviation $10\%$. Lastly, scenario $3$ depicts the most extreme case with rare but very severe jumps with on average one jump every twelve and a half years of mean $-20\%$ and standard deviation $15\%$. 

Moreover, parameter values for the log-AR(1) implied volatility model are kept fixed for all three scenarios and are presented in \cref{table:all_OU_params}. Note that the long-run parameter $\vartheta$ is set at the logarithm of the yearly standard deviation of log-returns with $\vartheta = \log 0.15$, and other parameters are chosen in an ad hoc fashion so as to produce reasonable values for implied volatilities. 

\begin{table} [ht]
	\caption {Parameters of the log-AR(1) model for the evolution of implied volatilities.}
	\label{table:all_OU_params}
	\begin{adjustwidth}{-1in}{-1in} 
		\centering
		\begin{tabular}{cccc}
			\hline
			$\kappa$ & $\vartheta $ & $\sigma_{\text{IV}}$ & $\varrho$
			\\
			\hline\noalign{\medskip}
			%
			$0.15$  &  $\log(0.15)$ & $0.06$ & $-0.6$
			\\    
			\noalign{\medskip}\hline
		\end{tabular}%
	\end{adjustwidth}
\end{table}



\subsubsection{Benchmarking results in the presence of jump risk}
\label{subsubsec:jump_risk}
\cref{table:ERP_one_year_put_MJD_analysis}
presents equal risk prices $C_0^{\star}$ and residual hedging risk $\epsilon^{\star}$ across the three scenarios of jump parameters and different trading instruments. 
\begin{table}
	\caption {Sensitivity analysis of equal risk prices $C_{0}^{\star}$ and residual hedging risk $\epsilon^{\star}$ to jump risk for OTM ($K=90$), ATM ($K=100$) and ITM ($K=110$) put options of maturity $T=1$.} \label{table:ERP_one_year_put_MJD_analysis}
	\renewcommand{\arraystretch}{1.15}
	\begin{adjustwidth}{-1in}{-1in} 
		\centering
		\begin{tabular}{lccccccccccc}
			\hline\noalign{\smallskip}
			& \multicolumn{3}{c}{$\text{OTM}$} & & \multicolumn{3}{c}{$\text{ATM}$} & & \multicolumn{3}{c}{$\text{ITM}$} \\
			\cline{2-4}\cline{6-8}\cline{10-12} Jump Scenario & $(1)$ & $(2)$ & $(3)$ &   & $(1)$ & $(2)$ & $(3)$ &   & $(1)$ & $(2)$ & $(3)$ \\
			%
			\hline\noalign{\medskip} 
			\multicolumn{4}{l}{\underline{\emph{$C_{0}^{\star}$}}} \\\noalign{\smallskip} 
			$\text{Daily stock}$        & $1.89$   & $2.58$     & $3.36$   &  & $5.21$   & $6.01$      & $6.81$    & & $10.81$   & $11.68$     & $12.13$  \\
			$\text{Monthly stock}$      & $1.97$   & $2.60$     & $3.31$   &  & $5.04$   & $5.77$      & $6.38$    & & $10.73$   & $11.44$     & $11.86$  \\
			$\text{1-month options}$    & $1.82$   & $2.24$     & $2.55$   &  & $4.99$   & $5.36$      & $5.60$    & & $10.48$   & $10.86$     & $10.83$  \\
			$\text{3-months options}$   & $1.74$   & $2.08$     & $2.39$   &  & $4.87$   & $5.12$      & $5.28$    & & $10.43$   & $10.51$     & $10.57$  \\
			&   &   &  & & & & & & & & \\
			\multicolumn{4}{l}{\underline{\emph{$\epsilon^{\star}$}}} \\\noalign{\smallskip} 
			$\text{Daily stock}$        & $1.09$   & $1.98$     & $2.67$   &  & $1.76$   & $2.74$      & $3.54$    & & $1.82$   & $2.78$     & $3.27$  \\
			$\text{Monthly stock}$      & $1.82$   & $2.52$     & $3.26$   &  & $3.00$   & $3.88$      & $4.57$    & & $3.07$   & $3.91$     & $4.37$  \\
			$\text{1-month options}$    & $0.76$   & $1.18$     & $1.52$   &  & $1.14$   & $1.53$      & $1.78$    & & $1.17$   & $1.56$     & $1.54$  \\
			$\text{3-months options}$   & $1.03$   & $1.37$     & $1.68$   &  & $1.59$   & $1.82$      & $2.02$    & & $1.70$   & $1.79$     & $1.88$  \\
			&   &   &  & & & & & & & & \\
			\multicolumn{4}{l}{\underline{\emph{$\epsilon^{\star}/C_{0}^{\star}$}}} \\\noalign{\smallskip} 
			$\text{Daily stock}$        & $0.58$   & $0.77$     & $0.79$   &  & $0.34$   & $0.46$      & $0.52$    & & $0.17$   & $0.24$     & $0.27$  \\
			$\text{Monthly stock}$      & $0.92$   & $0.97$     & $0.99$   &  & $0.60$   & $0.67$      & $0.72$    & & $0.29$   & $0.34$     & $0.37$  \\
			$\text{1-month options}$    & $0.42$   & $0.53$     & $0.60$   &  & $0.23$   & $0.28$      & $0.32$    & & $0.11$   & $0.14$     & $0.14$  \\
			$\text{3-months options}$   & $0.59$   & $0.66$     & $0.70$   &  & $0.33$   & $0.36$      & $0.38$    & & $0.16$   & $0.17$     & $0.18$  \\
			\noalign{\medskip}\hline
		\end{tabular}%
	\end{adjustwidth}
	Notes: Results are computed based on $100,\!000$ independent paths generated from the Merton Jump-Diffusion model for the underlying (see \cref{subsubsec:MJD_model} for model description). Three different sets of parameters values are considered with $\lambda = \{1, 0.25, 0.08\}$, $\mu_J = \{-0.05, -0.10, -0.20\}$ and $\sigma_J = \{0.05, 0.10, 0.15\}$ respectively for jump scenario $1$, $2$, and $3$ (see \cref{table:all_MJD} for all parameters values). \textit{Hedging instruments}: daily or monthly rebalancing with the underlying stock and 1-month or 3-months options with ATM calls and puts. Options used as hedging instruments are priced with implied volatility modeled with a log-AR(1) dynamics (see \cref{subsec:dynamics_IV} for model description and \cref{table:all_OU_params} for parameters values). The training of neural networks is done as described in \cref{subsubsec:hyperparameters}. The confidence level of the CVaR measure is $\alpha = 0.95$.
\end{table}
%
Numerical values indicate that in the presence of jump risk, hedging with options entails significant reduction of both equal risk prices and market incompleteness as compared to hedging solely with the underlying stock across all moneyness levels and jump risk scenarios. The reduction in hedging residual risk by trading options is obtained despite less frequent rebalancing than when only the stock is used. 
These results add additional evidence that options are indeed non-redundant as prescribed by the Black-Scholes world: 
the equal risk pricing framework dictates that hedging with options in the presence of jump risk can significantly impact both derivative premiums and hedging risk as quantified by our incompleteness metrics. 

The relative reduction achieved in $C_0^{\star}$ with 1-month and 3-months options as compared to hedging with the stock is most important for OTM puts, followed by ATM and ITM contracts. For instance, the relative reduction obtained with 3-months options hedging over daily stock hedging ranges across the three jump risk scenarios between $8\%$ to $29\%$ for OTM, $6\%$ to $22\%$ for ATM and $4\%$ to $13\%$ for ITM puts.\footnote{
	If $C_0^{\star}(\text{daily stock})$ and $C_0^{\star}(\text{3-months options})$ are equal risk prices obtained respectively by hedging with the stock on a daily basis and with 3-months options, the relative reduction is computed as $1 - \frac{C_0^{\star}(\text{3-months options})}{C_0^{\star}(\text{daily stock})}$ for all examples. 
} 
This reduction in $C_0^{\star}$ when using options as hedging instruments can be explained by the following observations. As pointed out in \cite{carbonneau2020equal}, the fact that a put option payoff is bounded below at zero entails that the short position hedging error has a thicker right tail than the long position hedging error. Also, it is widely documented in the literature that hedging jump risk with options significantly dampens tail risk as compared to using only the underlying stock (see for instance \cite{coleman2007robustly} and \cite{carbonneau2020deep}).\footnote{
\cite{horvath2021deep} deep hedge derivatives under a rough Bergomi volatility model by trading the underlying stock and a variance swap. The latter paper shows that this dynamics exhibits jump-like behaviour when discretized. As results presented in this current paper highlights the fact that global hedging jump risk with option hedges is very effective, deep hedging with options could also potentially be effective under such rough volatility models.
}
Consequently, 
the choice of trading options to mitigate jump risk reduces the measured risk exposure of both the long and short positions, but the thicker right tail for the short position hedging error entails a larger decrease for the latter than for the long position. In such situations, the ERP framework dictates that the long position should be compensated with a lower derivative premium $C_0^{\star}$ to equalize residual hedging risk of both positions.

Moreover, values for both $\epsilon^{\star}$-metrics indicate that in the presence of jump risk, the use of options contributes significantly to the reduction of market incompleteness as both the long and short position hedges achieve risk reduction when compared to trading only with the stock. 
The latter conclusion is in itself not novel, and is widely documented in the literature (see, for instance, \cite{tankov2003financial} and the many references therein). Indeed, this is a consequence of the well-known convex property of put option prices, which implies that hedging random jumps solely with the underlying stock is ineffective. Our $\epsilon^{\star}$-metrics have the advantage of allowing for a precise quantification of such reduction in residual hedging risk achieved through the use of options as hedging instruments. 

The sensitivity of equal risk prices and residual hedging risk across the three jump risk scenarios for each set of hedging instruments is now examined.
Numerical results presented in \cref{table:ERP_one_year_put_MJD_analysis} indicate that for a fixed set of hedging instruments, both the equal risk price and the level of incompleteness increases with the severity of jumps across all moneyness levels.
Indeed, the relative increase of equal risk prices observed under scenario $3$ as compared to scenario $1$ respectively for OTM, ATM and ITM puts is $78\%, 31\%$ and $12\%$ with the daily stock, $68\%$, $27\%$ and $11\%$ with the monthly stock, $40\%$, $12\%$ and $3\%$ with 1-month options and $38\%$, $8\%$ and $1\%$ with 3-months options.\footnote{
	For a fixed hedging instrument and moneyness level, if $C_0^{\star}(\text{scenario 1})$ and $C_0^{\star}(\text{scenario 3})$ are respectively the equal risk price obtained under jump risk scenario $1$ and $3$, the relative increase is computed as $\frac{C_0^{\star}(\text{scenario 3})}{C_0^{\star}(\text{scenario 1})} - 1$.
} Similar observations can be made for both incompleteness metrics: increases in jump severity leads to larger $\epsilon^{\star}$ and $\epsilon^{\star}/C_0^{\star}$. This positive association between both equal risk prices and the level of market incompleteness to jump severity can be explained by the following observations. For a fixed hedging instrument and moneyness level, the long measured risk exposure is closed to invariant to jump severity (i.e. similar values across the three jump risk scenarios). The latter stems from the fact that jump dynamics considered in this paper predominantly entail negative jumps, which result in a thicker left tail for the long position hedging error (i.e. hedging gains) as jump severity increases, but in close to no impact on the right tail of the long position hedging error. In contrast, since the right tail weight of the short position hedging error increases with the expected (negative) magnitude and volatility of jumps, the short measured risk exposure always increases going from scenario $1$ to scenario $3$, which consequently increases both the equal risk price and the level of market incompleteness.


\subsection{Sensitivity of equal risk pricing to volatility risk}
\label{subsec:sensitivity_GARCH}
Having examined the impact of jump risk on the ERP framework, the impact of volatility risk is now studied.
In the same spirit as analyses done for jump risk, three different sets of parameters are considered for the GARCH dynamics which imply annualized stationary (expected) volatilities of $10\%, 15\%$ and $20\%$.\footnote{
	The annualized stationary volatility with $252$ days per year is computed as 
	$$\sqrt{\frac{252\omega}{1 - \upsilon (1+\gamma^{2})-\beta}}.$$
} 
The three sets of parameters are presented in \cref{table:all_GARCH}. 
Note that every parameter is fixed for all three sets, except for the level parameter $\omega$, which is adjusted to attain the wanted stationary volatility. The value of the drift parameter $\mu$ is set such that the yearly expected value of log-returns is $10\%$. Also, values for $\{\upsilon , \gamma, \beta\}$ are inspired from parameters estimated with maximum likelihood on a time series of daily log-returns on the S\&P 500 index for the period 1986-12-31 to 2010-04-01 used in \cite{carbonneau2020equal}. The same setup is considered as in \cref{subsec:sensitivity_MJD} in terms of the derivative to be priced ($1$-year maturity European puts) and for the choice of hedging instruments (underlying stock traded on a daily or monthly basis and $1$-month or $3$-months maturities ATM calls and puts).
The same parameters as in the study of jump risk conducted in \cref{subsec:sensitivity_MJD} are used for $\{\kappa, \sigma_{\text{IV}}, \varrho\}$ of the log-AR(1) dynamics for the evolution of $1$-month and $3$-months ATM IVs (i.e. $\kappa = 0.15, \sigma_{\text{IV}}=0.06$ and $\varrho = -0.6$), except for the long-run parameter $\vartheta$, which is set to be in line with the underlying GARCH process as $\log(0.10), \log(0.15)$ and $\log(0.20)$ when the stationary volatility is $10\%$, $15\%$ and $20\%$, respectively. It is worth highlighting that the choice of modeling implied volatilities for short-term options with higher and smaller average levels enables us to assess the impact of larger and smaller average costs for trading options on the equal risk price and residual hedging risk of longer-term options.
\begin{table} [ht]
	\caption {Parameters of the GJR-GARCH model for $10\%, 15\%$ and $20\%$ stationary yearly volatilities.} \label{table:all_GARCH}
	\begin{adjustwidth}{-1in}{-1in} 
		\centering  
		\begin{tabular}{cccccc}
			\hline
			$\text{Stationary volatility}$ & $\mu$ & $\omega$ & $\upsilon$ & $\gamma$ & $\beta$
			\\
			\hline\noalign{\medskip}
			$\text{10\%}$ & $3.968\text{e-}04$  &  $8.730\text{e-}07$ & $0.05$ & $0.6$ & $0.91$
			\\
			%
			$\text{15\%}$ & $3.968\text{e-}04$  &  $1.964\text{e-}06$ & $0.05$ & $0.6$ & $0.91$
			\\
			%
			$\text{20\%}$ & $3.968\text{e-}04$  &  $3.492\text{e-}06$ & $0.05$ & $0.6$ & $0.91$
			\\    
			\noalign{\medskip}\hline
		\end{tabular}%
	\end{adjustwidth}
\end{table}

\subsubsection{Benchmarking results with volatility risk}
\cref{table:ERP_one_year_put_vol_analysis} presents equal risk prices $C_0^{\star}$ and $\epsilon^{\star}$-metrics for put options of $1$ year maturity across the three sets of volatility risk parameters and hedging instruments. 
\begin{table}
	\caption {Sensitivity analysis of equal risk prices $C_{0}^{\star}$ and residual hedging risk $\epsilon^{\star}$ to volatility risk for OTM ($K=90$), ATM ($K=100$) and ITM ($K=110$) put options of maturity $T=1$.} \label{table:ERP_one_year_put_vol_analysis}
	\renewcommand{\arraystretch}{1.15}
	\begin{adjustwidth}{-1in}{-1in} 
		\centering
		\begin{tabular}{lccccccccccc}
			\hline\noalign{\smallskip}
			& \multicolumn{3}{c}{$\text{OTM}$} & & \multicolumn{3}{c}{$\text{ATM}$} & & \multicolumn{3}{c}{$\text{ITM}$} \\
			\cline{2-4}\cline{6-8}\cline{10-12} Stationary volatility   & $10\%$ & $15\%$ & $20\%$ &   & $10\%$ & $15\%$ & $20\%$ &   & $10\%$ & $15\%$ & $20\%$ \\
			\hline\noalign{\medskip} 
			\multicolumn{4}{l}{\underline{\emph{$C_{0}^{\star}$}}} \\\noalign{\smallskip} 
			$\text{Daily stock}$        & $1.01$   & $2.35$     & $3.85$   &  & $3.23$   & $5.36$      & $7.24$    & & $8.56$   & $10.55$     & $12.51$  \\
			$\text{Monthly stock}$      & $1.17$   & $2.65$     & $4.23$   &  & $3.37$   & $5.44$      & $7.58$    & & $8.85$   & $10.82$     & $12.88$  \\
			$\text{1-month options}$    & $0.56$   & $1.74$     & $3.27$   &  & $2.86$   & $4.87$      & $6.89$    & & $8.46$   & $10.32$     & $12.35$  \\
			$\text{3-months options}$   & $0.76$   & $2.07$     & $3.65$   &  & $3.01$   & $5.08$      & $7.16$    & & $8.51$   & $10.38$     & $12.44$  \\
			&   &   &  & & & & & & & & \\
			\multicolumn{4}{l}{\underline{\emph{$\epsilon^{\star}$}}} \\\noalign{\smallskip} 
			$\text{Daily stock}$        & $0.77$   & $1.51$     & $2.21$   &  & $1.28$   & $2.12$      & $2.67$    & & $1.12$   & $1.98$     & $2.65$  \\
			$\text{Monthly stock}$      & $1.15$   & $2.44$     & $3.67$   &  & $2.15$   & $3.41$      & $4.62$    & & $1.92$   & $3.29$     & $4.56$  \\
			$\text{1-month options}$    & $0.26$   & $0.65$     & $1.06$   &  & $0.59$   & $1.00$      & $1.36$    & & $0.62$   & $1.03$     & $1.39$  \\
			$\text{3-months options}$   & $0.59$   & $1.32$     & $2.02$   &  & $1.10$   & $1.77$      & $2.41$    & & $1.04$   & $1.68$     & $2.31$  \\
			&   &   &  & & & & & & & & \\
			\multicolumn{4}{l}{\underline{\emph{$\epsilon^{\star}/C_{0}^{\star}$}}} \\\noalign{\smallskip} 
			$\text{Daily stock}$        & $0.77$   & $0.64$     & $0.57$   &  & $0.40$   & $0.40$      & $0.37$    & & $0.13$   & $0.19$     & $0.21$  \\
			$\text{Monthly stock}$      & $0.99$   & $0.92$     & $0.87$   &  & $0.64$   & $0.63$      & $0.61$    & & $0.22$   & $0.30$     & $0.35$  \\
			$\text{1-month options}$    & $0.45$   & $0.37$     & $0.32$   &  & $0.21$   & $0.20$      & $0.20$    & & $0.07$   & $0.10$     & $0.11$  \\
			$\text{3-months options}$   & $0.77$   & $0.64$     & $0.55$   &  & $0.36$   & $0.35$      & $0.34$    & & $0.12$   & $0.16$     & $0.19$  \\
			\noalign{\medskip}\hline
		\end{tabular}%
	\end{adjustwidth}
	Notes: Results are computed based on $100,\!000$ independent paths generated from the GJR-GARCH(1,1) model for the underlying with three sets of parameters implying stationary yearly volatilities of $10\%, 15\%$ and $20\%$ (see \cref{subsubsec:GARCH_model} for model description and \cref{table:all_GARCH} for parameters values).  \textit{Hedging instruments}: daily or monthly rebalancing with the underlying stock and 1-month or 3-months options with ATM calls and puts. Options used as hedging instruments are priced with implied volatility modeled as a log-AR(1) dynamics with $\kappa = 0.15, \sigma_{\text{IV}} = 0.06$ and $\varrho = -0.6$ for all cases, and $\vartheta$ set to $\log(0.10), \log(0.15)$ and $\log(0.20)$ when the GARCH stationary volatility is $10\%, 15\%$ and $20\%$, respectively (see \cref{subsec:dynamics_IV} for the log-AR(1) model description). The training of neural networks is done as described in \cref{subsubsec:hyperparameters}. The confidence level of the CVaR measure is $\alpha = 0.95$.
\end{table}
Numerical results indicate that in the presence of volatility risk, the use of options as hedging instruments can reduce $C_0^{\star}$ as compared to daily stock hedging. However, this impact on $C_0^{\star}$ when trading options can be marginal and is highly sensitive to the moneyness level of the put option being priced as well as to the maturity of the traded options. Furthermore, the impact on $C_0^{\star}$ of the use of options within hedges tends to diminish when traded options are more costly (i.e. as the average level of implied and GARCH volatility increases). Indeed, the relative reduction in equal risk prices achieved with 1-month options hedging as compared to daily stock hedging with $10\%, 15\%$ and $20\%$ stationary volatility is respectively $44\%, 26\%$ and $15\%$ for OTM puts, $12\%, 9\%$ and $5\%$ for ATM and $1\%, 2\%$ and $1\%$ for ITM options.\footnote{
	If $C_0^{\star}(\text{daily stock})$ and $C_0^{\star}(\text{1-month options})$ are respectively the equal risk price obtained by hedging with the stock on a daily basis and with 1-month options, the relative reduction is computed as $1 - \frac{C_0^{\star}(\text{1-month options})}{C_0^{\star}(\text{daily stock})}$ for all examples. 
} However, the relative reduction in $C_0^{\star}$ with $3$-months option hedges as compared to using the stock on a daily basis is overall much more marginal, with the notable exceptions of OTM and ATM puts with $10\%$ stationary volatility which achieve respectively $25\%$ and $7\%$ reduction as well as for the OTM moneyness under $15\%$ stationary volatility with a $12\%$ reduction. 
Also, as expected, values presented in \cref{table:ERP_one_year_put_vol_analysis} confirm that the level of market incompleteness as measured by the $\epsilon^{\star}$ metric has a positive relationship with the average level of stationary volatility for all hedging instruments.

The previously described observations about the impact of option hedges on both equal risk prices and residual hedging risk all stem from the realized reduction in measured risk exposure by the long and short positions. However, contrarily to results obtained with jump risk, the reduction in measured risk exposure when hedging volatility risk with options can be very similar for both the long and short positions, whereas with jump risk, the reduction is asymmetric by always favoring the short position with a larger reduction. The latter can be explained by the fact that volatility risk impacts both upside and downside risk, while the impact of jump risk dynamics considered in this paper is very asymmetric by entailing significantly more weight on the right (resp. left) tail of the short (resp. long) hedging error with predominantly negative jumps. 
Values presented in \cref{table:ERP_one_year_put_vol_analysis} confirm this analysis of the interrelation between volatility risk and the choice of hedging instruments. For instance, for ITM puts, the measured risk exposure of the long and short positions decreases by a similar amount when trading $1$-month or $3$-months options as compared to daily stock hedges, which explains the significant decrease in $\epsilon^{\star}$, but also the insensitivity of $C_0^{\star}$ to the choice of hedging instruments and rebalancing frequency. On the other hand, for OTM puts, 1-month and 3-months option hedges results in larger decreases of measured risk exposure for the short position than for the long position, which explains the reduction in $C_0^{\star}$ and $\epsilon^{\star}$ as compared to daily stock hedges. 

Lastly, it is very interesting to observe that the average price level of short-term options used as hedging instruments is effectively reflected into the equal risk price of longer-term options. Indeed, numerical results for $C_0^{\star}$ presented in \cref{table:ERP_one_year_put_vol_analysis} highlight the fact that higher hedging options implied volatilities for $1$-month and $3$-months ATM calls and puts leads to higher equal risk prices for $1$-year maturity puts. Furthermore, to isolate the idiosyncratic contribution of the variations of option prices used as hedging instruments on the equal risk price from the impact of the stationarity volatility of the GARCH process, the authors also tested fixing the stationarity volatility of the GARCH process to $15\%$ and setting the long-run parameter of the IV process to $14\%$ and $16\%$. These results presented in the Supplementary Material, Table SM2, confirm that higher implied volatilities for options used as hedging instruments leads to higher equal risk prices.
All of these benchmarking results demonstrate the potential of the ERP framework as a fair valuation approach consistent with observable market prices. For instance, the ERP framework could be used to price and optimally hedge over-the-counter derivatives with vanilla options. An additional potential application is the marking-to-market of less liquid long-term derivatives (e.g. Long-Term Equity AnticiPation Securities (LEAPS)) consistently with highly liquid shorter-term option hedges. The ERP framework could also be used for the fair valuation of segregated funds guarantees, which are equivalent to very long-term (up to 40 years) derivatives sold by insurers.\footnote{
	Note that \cite{carbonneau2020deep} demonstrates the potential of the deep hedging algorithm for global hedging long-term lookback options embedded in segregated funds guarantees with multiple hedging instruments. It is also worth highlighting that \cite{barigou2020insurance} developed a pricing scheme consistent with local non-quadratic hedging procedures for insurance liabilities which relies on neural networks.
	} 
Indeed, International Financial Reporting Standards 17 (IFRS 17, \cite{IFRS17}) mandates a market consistent valuation of options embedded in segregated funds guarantees with readily available observable market prices at the measurement date. The ERP framework could potentially be applied to price such very long-term options consistently with shorter-term implied volatility surface dynamics, with the latter being much less challenging to calibrate due to the higher liquidity of short-term options.\footnote{
In the context of segregated funds, the short position of the embedded option is assumed to be held by an insurance company who has to provide a quote and mitigate its risk exposure. The long position is held by an unsophisticated investor who will not be hedging his risk exposure. Nevertheless, as IFRS 17 mandates the use of a fair valuation approach for embedded options consistent with observable market prices, the ERP framework could potentially be used in this context.
} 

	\subsection{Sensitivity analyses to the confidence level of $\text{CVaR}_{\alpha}$}
	\label{subsec:sensitivity_risk_aversion}
	This section conducts sensitivity analyses with respect to the choice of convex risk measure on the ERP framework when trading exclusively options. Similarly to the work of \cite{carbonneau2020equal}, values for equal risk prices and $\epsilon^{\star}$-metrics are examined across the confidence levels $0.90, 0.95$ and $0.99$ for the $\text{CVaR}_{\alpha}$ measure. As argued in the latter paper, higher confidence levels corresponds to more risk averse agents by concentrating more relative weight on losses of larger magnitude. The main finding of the sensitivity analysis conducted in \cite{carbonneau2020equal} is that when trading exclusively the underlying stock, higher confidence levels leads to larger values for $C_0^{\star}$ and $\epsilon^{\star}$ metrics. The objective of this section is to assess if this finding is robust to the use of short-term option hedges instead of the underlying stock. 
	For each confidence level, the authors of the current paper computed both equal risk prices and residual hedging risk obtained by trading $3$-months ATM calls and puts with the same setup as in \cref{subsec:sensitivity_MJD} and \cref{subsec:sensitivity_GARCH}, i.e. for all three jump and volatility scenarios of parameters.\footnote{
	Unreported tests performed by the authors show that values lower than $0.90$ for the confidence level of $\text{CVaR}_{\alpha}$ with $1$-month and $3$-months option hedges lead to trading policies with significantly larger tail risk in a way which would deem such policies as inadmissible by hedgers. Using the $\text{CVaR}_{0.90}$ measure with $1$-month options also resulted in trading policies with significantly larger tail risk. However, this large increase in tail risk was not observed with the $\text{CVaR}_{0.95}$ and $\text{CVaR}_{0.99}$ measures when trading $1$-month options, nor with $\text{CVaR}_{0.90}$, $\text{CVaR}_{0.95}$ and $\text{CVaR}_{0.99}$ when trading $3$-months options. These observations motivated the choice of performing sensitivity analysis for $\text{CVaR}_{\alpha}$ with $\alpha = 0.90, 0.95$ and $0.99$ exclusively when trading $3$-months options.
	} Overall, the main conclusions are found to be qualitatively similar for all of the different setups. Thus, to save space, values for equal risk prices and residual hedging risk are only reported under the MJD dynamics with jump risk scenario $2$ by trading $3$-months options; these results are presented in \cref{table:sensitivity_analysis_CVAR_090_095_099}. The interested reader in numerical results obtained under jump risk scenarios $1$ and $3$ as well as under the three sets of volatility risk parameters is referred to Table SM3 and Table SM4 of the Supplementary Material. 
	
	\begin{table} [ht]
		\caption {Sensitivity analysis of equal risk prices $C_{0}^{\star}$ and residual hedging risk $\epsilon^{\star}$ for OTM ($K=90$), ATM ($K=100$) and ITM ($K=110$) put options of maturity $T=1$ under the MJD dynamics with jump risk scenario $2$.} \label{table:sensitivity_analysis_CVAR_090_095_099}
		\renewcommand{\arraystretch}{1.15}
		\begin{adjustwidth}{-1in}{-1in} 
			\centering
			\begin{tabular}{lccccccccccc}
				\hline\noalign{\smallskip}
				& \multicolumn{3}{c}{$C_{0}^{\star}$} & & \multicolumn{3}{c}{$\epsilon^{\star}$} & & \multicolumn{3}{c}{$\epsilon^{\star}/C_{0}^{\star}$} \\
				\cline{2-4}\cline{6-8}\cline{10-12}   Moneyness   & $\text{OTM}$ & $\text{ATM}$ & $\text{ITM}$ &   & $\text{OTM}$ & $\text{ATM}$ & $\text{ITM}$ &   & $\text{OTM}$ & $\text{ATM}$ & $\text{ITM}$ \\
				\hline\noalign{\medskip} 
				%
				$\text{CVaR}_{0.90}$ & $1.86$   & $4.93$  & $10.40$  &   & $0.99$   & $1.43$  & $1.50$  &   & $0.53$  & $0.29$  & $0.14$ \\
				$\text{CVaR}_{0.95}$ & $12\%$   & $4\%$   & $1\%$    &   & $39\%$   & $28\%$  & $20\%$  &   & $24\%$  & $23\%$  & $18\%$ \\
				$\text{CVaR}_{0.99}$ & $40\%$   & $10\%$  & $4\%$    &   & $116\%$  & $76\%$  & $65\%$  &   & $54\%$  & $60\%$  & $59\%$ \\
				\noalign{\medskip}\hline
			\end{tabular}%
		\end{adjustwidth}
		Notes: Results are computed based on $100,\!000$ independent paths generated from the Merton Jump-Diffusion model for the underlying (see \cref{subsubsec:MJD_model} for model description) with parameters $\nu = 0.1111, \sigma = 0.1323, \lambda = 0.25, \mu_J = -0.10$ and $\sigma_J = 0.10$ corresponding to jump risk scenario $2$ of \cref{table:all_MJD}. Hedging instruments consist of 3-months ATM calls and puts priced with implied volatility modeled with a log-AR(1) dynamics (see \cref{subsec:dynamics_IV} for model description and \cref{table:all_OU_params} for parameters values). The training of neural networks is done as described in \cref{subsubsec:hyperparameters}.  Values for the $\text{CVaR}_{0.95}$ and $\text{CVaR}_{0.99}$ measures are expressed relative to $\text{CVaR}_{0.90}$ (\% increase).
	\end{table}
	Numerical values reported in \cref{table:sensitivity_analysis_CVAR_090_095_099} indicate that with option hedges, an increase in the confidence level parameter of the $\text{CVaR}_{\alpha}$ measure leads to larger equal risk prices $C_0^{\star}$ and residual hedging risk $\epsilon^{\star}$ across all examples. These results confirm that the finding of \cite{carbonneau2020equal} with respect to the sensitivity of $C_0^{\star}$ and $\epsilon^{\star}$ to the risk aversion of the hedger is robust to using exclusively options as hedging instruments. Furthermore, values for equal risk prices $C_0^{\star}$ show a largest increase when using $\text{CVaR}_{0.95}$ and $\text{CVaR}_{0.99}$ as compared to $\text{CVaR}_{0.90}$ for OTM puts, followed by ATM and ITM moneyness levels; the same conclusion was observed in \cite{carbonneau2020equal} when trading the underlying stock. 
	The increase in $C_0^{\star}$ with the risk aversion level of the hedger stems from the thicker right tail of the short position hedging error than for the long position hedging error. The latter observation is consistent with previous analyses: while option hedges are more effective than stock hedges in the presence of equity jump risk as demonstrated in \cref{subsec:sensitivity_MJD}, their inclusion within hedging portfolios does not fully mitigate the asymmetry in tail risk of the residual hedging error.
	

	\subsection{Benchmarking of equal risk prices to variance-optimal premiums}
	\label{subsec:benchmark_QDH_ERP}	
 	This section presents the benchmarking of equal risk prices to derivative premiums obtained with \textit{variance-optimal hedging} procedures (VO, \cite{schweizer1995variance}), also commonly called \textit{global quadratic hedging}. Variance-optimal hedging solves jointly for the initial capital investment and a self-financing strategy minimizing the expected value of the squared hedging error:
 	\begin{align}
 	\underset{\delta\in \Pi, V_0 \in \mathbb{R}}{\min} \, \E \left[\left(\Phi(S_{N}^{(0,b)}) -B_{N}(V_0 + G_{N}^{\delta})\right)^{2}\right]. \label{eq:ref_QDH_optimization}
 	\end{align}
 	The optimized initial capital investment denoted hereafter as $C_0^{(VO)}$ can be viewed as the production cost of $\Phi$, since the resulting dynamic trading strategy replicates the derivative's payoff as closely as possible in a quadratic sense. The optimization problem \eqref{eq:ref_QDH_optimization} can also be solved in a similar fashion as the non-quadratic global hedging problems embedded in the ERP framework, but with two distinctions: the initial capital investment is treated as an additional trainable parameter and a single neural network is considered since the optimal trading strategy is the same for the long and short position due to the quadratic penalty.\footnote{
 	\cite{cao2020discrete} showed that the deep hedging algorithm for variance-optimal hedging problems provides good approximations of optimal initial capital investments by comparing the optimized values to known formulas.
 	} The reader is referred to \cref{appendix:QDH_price} for a complete description of the numerical scheme for variance-optimal hedging implemented in this study. 
 	
 	The setup considered for the examination of this benchmarking is the same as in \cref{subsec:sensitivity_MJD} with the MJD dynamics under the three jump risk scenarios, with the exception of the confidence level of the $\text{CVaR}_{\alpha}$ measure, which is studied at first with $\alpha = 0.95$ fixed as in \cref{subsec:sensitivity_MJD} and \cref{subsec:sensitivity_GARCH}, and subsequently across $\alpha = 0.90, 0.95$ and $0.99$ as in \cref{subsec:sensitivity_risk_aversion}. Note that the authors also conducted the same experiments under the setup of \cref{subsec:sensitivity_GARCH} with volatility risk, and found that the main qualitative conclusions are very similar. The reader is referred to Table SM$6$ and Table SM$8$ of the Supplementary Material for the benchmarking of ERP to VO procedures in the presence of volatility risk.
 	
 	 
 	\subsubsection{Benchmarking results}
 	\cref{table:ERP_RN_prices_one_year_JUMPS} presents benchmarking results of equal risk prices $C_0^{\star}$ to variance-optimal prices $C_0^{(VO)}$ under the MJD dynamics with the $\text{CVaR}_{0.95}$ measure.
 	\begin{table} [ht]
 		\caption {Equal risk prices $C_{0}^{\star}$ and variance-optimal (VO) prices $C_0^{(VO)}$ with jump risk for OTM ($K=90$), ATM ($K=100$) and ITM ($K=110$) put options of maturity $T=1$.} \label{table:ERP_RN_prices_one_year_JUMPS}
 		\renewcommand{\arraystretch}{1.15}
 		\begin{adjustwidth}{-1in}{-1in} 
 			\centering
 			\begin{tabular}{lccccccccccc}
 				\hline\noalign{\smallskip}
 				& \multicolumn{3}{c}{$\text{OTM}$} & & \multicolumn{3}{c}{$\text{ATM}$} & & \multicolumn{3}{c}{$\text{ITM}$} \\
 				\cline{2-4}\cline{6-8}\cline{10-12} Jump Scenario & $(1)$ & $(2)$ & $(3)$ &   & $(1)$ & $(2)$ & $(3)$ &   & $(1)$ & $(2)$ & $(3)$ \\	
 				\hline\noalign{\medskip} 
 				\multicolumn{4}{l}{\underline{\emph{$C_0^{(VO)}$}}} \\\noalign{\smallskip} 
 				%
 				$\text{Daily stock}$        & $1.62$   & $1.77$     & $1.92$   &  & $4.71$   & $4.79$     & $4.80$    & & $10.20$   & $10.18$     & $10.12$  \\
				$\text{Monthly stock}$      & $1.55$   & $1.73$     & $1.86$   &  & $4.62$   & $4.72$     & $4.72$    & & $10.14$   & $10.11$     & $10.05$  \\
				$\text{1-month options}$    & $1.71$   & $2.04$     & $2.38$   &  & $4.82$   & $5.11$     & $5.31$    & & $10.27$   & $10.45$     & $10.52$  \\
				$\text{3-months options}$   & $1.58$   & $1.83$     & $2.08$   &  & $4.64$   & $4.82$     & $4.97$    & & $10.11$   & $10.15$     & $10.21$  \\
 				&   &   &  & & & & & & & & \\
 				\multicolumn{4}{l}{\underline{\emph{$C_{0}^{\star}$}}} \\\noalign{\smallskip} 
 				$\text{Daily stock}$        & $17\%$   & $45\%$     & $75\%$   &  & $11\%$  & $25\%$    & $42\%$   & & $6\%$   & $15\%$    & $20\%$  \\
 				$\text{Monthly stock}$      & $27\%$   & $51\%$     & $78\%$   &  & $9\%$   & $22\%$    & $35\%$   & & $6\%$   & $13\%$    & $18\%$  \\
 				$\text{1-month options}$    & $6\%$    & $10\%$     & $7\%$    &  & $3\%$   & $5\%$     & $6\%$    & & $2\%$   & $4\%$     & $3\%$  \\
 				$\text{3-months options}$   & $10\%$   & $14\%$     & $15\%$   &  & $5\%$   & $6\%$     & $6\%$    & & $3\%$   & $4\%$     & $4\%$  \\
 				\noalign{\medskip}\hline
 			\end{tabular}%
 		\end{adjustwidth}
		Notes: Results are computed based on $100,\!000$ independent paths generated from the Merton Jump-Diffusion model for the underlying (see \cref{subsubsec:MJD_model} for model description). Three different sets of parameters values are considered with $\lambda = \{1, 0.25, 0.08\}$, $\mu_J = \{-0.05, -0.10, -0.20\}$ and $\sigma_J = \{0.05, 0.10, 0.15\}$ respectively for the jump scenario $1$, $2$, and $3$ (see \cref{table:all_MJD} for all parameters values).  \textit{Hedging instruments}: daily or monthly rebalancing with the underlying stock and 1-month or 3-months options with ATM calls and puts. Options used as hedging instruments are priced with implied volatility modeled with a log-AR(1) dynamics (see \cref{subsec:dynamics_IV} for model description and \cref{table:all_OU_params} for parameters values). The training of neural networks for ERP and VO hedging is done as described in \cref{subsubsec:hyperparameters} and \cref{appendix:QDH_price}, respectively. The confidence level of the CVaR measure is $\alpha = 0.95$. $C_0^{\star}$ are expressed relative to $C_0^{(VO)}$ (\% increase).
 	\end{table}
	Numerical experiments show that $C_0^{\star}$ is at least larger than $C_0^{(VO)}$ for all examples, but the relative increase is always smaller and less sensitive to jump severity when trading options. Furthermore, the relative increase in derivative premiums observed with the ERP framework over VO hedging is the largest for OTM puts, followed by ATM and ITM options across all jump risk scenarios and hedging instruments. For instance, the relative increase in $C_0^{\star}$ as compared to $C_0^{(VO)}$ when trading the daily stock ranges from jump scenario $1$ to scenario $3$ between $17\%$ to $75\%$ for OTM puts, $11\%$ to $42\%$ for ATM and $6\%$ to $20\%$ for ITM options.\footnote{
		The relative increase is computed as $\frac{C_0^{\star}}{C_0^{(VO)}}-1$ for all examples. 
	} On the other hand, the relative increase in $C_0^{\star}$ as compared to $C_0^{(VO)}$ is much less sensitive to jump severity when trading 1-month options by ranging from scenario $1$ to scenario $3$ between $6\%$ to $10\%$ for OTM puts, $3\%$ to $6\%$ for ATM and $2\%$ to $4\%$ for ITM. Based on these results, we can assert that although both derivative valuation schemes are consistent with optimal trading criteria, the choice of hedging instrument and pricing procedure (hence implicitly of the treatment of hedging gains and losses) has a material impact on resulting derivative premiums and must thus be carefully chosen.

    This smaller disparity between equal risk and variance-optimal prices with option hedges is in line with previous analyses: in the presence of jump or volatility risk, hedging with options entails significant reduction of the level market incompleteness as compared to trading solely the underlying stock. In such cases, premiums obtained with both derivative valuation approaches should be closer with the limiting case of being the same in a complete market.\footnote{
    	To further illustrate this phenomenon, the authors also performed the same benchmarking with the Black-Scholes dynamics under which market incompleteness solely stems from discrete-time trading. The latter results are presented in the Supplementary Material. Numerical values show that under the Black-Scholes dynamics, trading the underlying stock on a daily basis leads for most combinations of moneyness level and yearly volatility to the closest derivative premiums between ERP and VO procedures as compared to the other hedging instruments (see Table SM5). Also, as expected under the Black-Scholes dynamics, daily stock hedging entails the smallest level of residual hedging risk across the different hedging instruments (see Table SM1).
    	} These observations expand upon the work of \cite{carbonneau2020equal}, which shows that equal risk prices of puts obtained by hedging solely with the underlying stock are always larger than risk-neutral prices computed under convential change of measures. Indeed, benchmarking results presented in this current paper provide important novel insights into this price inflation phenomenon observed with the ERP framework: the disparity between equal risk and variance-optimal prices is always significantly smaller and less sensitive to stylized features of risky assets (e.g. jump or volatility risk) when option hedges are considered instead of trading exclusively the underlying stock. 
	
	Moreover, \cref{table:ERP_vs_VO_sensitivity_alpha_JUMP} presents benchmarking results of $C_0^{\star}$ to $C_0^{(VO)}$ with $\text{CVaR}_{0.90}, \text{CVaR}_{0.95}$ and $\text{CVaR}_{0.99}$ measures with $3$-months option hedges.
	\begin{table}
	\caption {Sensitivity analysis of equal risk prices $C_{0}^{\star}$ with $\text{CVaR}_{0.90}, \text{CVaR}_{0.95}$ and $\text{CVaR}_{0.99}$ measures to variance-optimal (VO) prices $C_0^{(VO)}$ under jump risk for OTM ($K=90$), ATM ($K=100$) and ITM ($K=110$) put options of maturity $T=1$.} \label{table:ERP_vs_VO_sensitivity_alpha_JUMP}
		\renewcommand{\arraystretch}{1.15}
		\begin{adjustwidth}{-1in}{-1in} 
			\centering
			\begin{tabular}{lccccccccccc}
				\hline\noalign{\smallskip}
				& \multicolumn{3}{c}{$\text{OTM}$} & & \multicolumn{3}{c}{$\text{ATM}$} & & \multicolumn{3}{c}{$\text{ITM}$} \\
				\cline{2-4}\cline{6-8}\cline{10-12} Jump Scenario & $(1)$ & $(2)$ & $(3)$ &   & $(1)$ & $(2)$ & $(3)$ &   & $(1)$ & $(2)$ & $(3)$ \\
				%
				\hline\noalign{\medskip} 
				%
				%
				$C_0^{(VO)}$  & $1.58$   & $1.83$     & $2.08$   &  & $4.64$   & $4.82$     & $4.97$    & & $10.11$   & $10.15$     & $10.21$  \\
				%
				$C_0^{\star}(\text{CVaR}_{0.90})$  & $3\%$    & $2\%$      & $0\%$   &  & $2\%$  & $2\%$     & $2\%$    & & $2\%$   & $2\%$     & $2\%$  \\
				$C_0^{\star}(\text{CVaR}_{0.95})$  & $10\%$   & $14\%$     & $15\%$   &  & $5\%$   & $6\%$     & $6\%$    & & $3\%$   & $4\%$     & $4\%$  \\
				$C_0^{\star}(\text{CVaR}_{0.99})$  & $32\%$   & $43\%$     & $55\%$   &  & $10\%$  & $12\%$    & $16\%$   & & $6\%$   & $6\%$     & $8\%$  \\
				\noalign{\medskip}\hline
			\end{tabular}%
		\end{adjustwidth}
		Notes: Results are computed based on $100,\!000$ independent paths generated from the Merton Jump-Diffusion model for the underlying (see \cref{subsubsec:MJD_model} for model description). Three different sets of parameters values are considered with $\lambda = \{1, 0.25, 0.08\}$, $\mu_J = \{-0.05, -0.10, -0.20\}$ and $\sigma_J = \{0.05, 0.10, 0.15\}$ respectively for jump scenario $1$, $2$, and $3$ (see \cref{table:all_MJD} for all parameters values). Hedging instruments consist of 3-months ATM calls and puts priced with implied volatility modeled as a log-AR(1) dynamics (see \cref{subsec:dynamics_IV} for model description and \cref{table:all_OU_params} for parameters values). The training of neural networks for ERP and VO hedging is done as described in \cref{subsubsec:hyperparameters} and \cref{appendix:QDH_price}, respectively. $C_0^{\star}$ with $\text{CVaR}_{0.90}, \text{CVaR}_{0.95}$ and $\text{CVaR}_{0.99}$ are expressed relative to $C_0^{(VO)}$ (\% increase).  
	\end{table}
	Values presented in this benchmarking demonstrate the ability of ERP, through the choice of convex risk measures, to span a large interval of prices which is close to encompass the variance-optimal premium. Indeed, under the $\text{CVaR}_{0.90}$ measure, we observe that $C_0^{\star}$ values are very close to $C_0^{(VO)}$ where the relative difference ranges between $0\%$ and $3\%$ across all moneynesses and jump risk scenarios. On the other hand, optimizing trading policies with more risk averse agents, i.e. with $\text{CVaR}_{0.95}$ or $\text{CVaR}_{0.99}$, provides a very wide range of derivative premiums with the ERP framework, especially for the OTM moneyness level. It is very interesting to note that this added flexibility of ERP procedures for pricing derivatives does not come at the expense of less effective hedging policies. Indeed, a major drawback of variance-optimal hedging lies in penalizing equally gains and losses through a quadratic penalty for hedging shortfalls. Conversely, the long and short trading policies solving the non-quadratic global hedging problems of the ERP framework are optimized to minimize a loss function which is possibly more in line with the financial objectives of the hedger by mainly (and most often exclusively) penalizing hedging losses, not gains. 
	
	\section{Conclusion}
	\label{section:conclusion}
	This paper studies the equal risk pricing (ERP) framework for pricing and hedging European derivatives in discrete-time with multiple hedging instruments. The ERP approach sets derivative prices as the value such that the optimally hedged residual risk of the long and short positions in the contingent claim are equal. The ERP setup of \cite{marzban2020equal} is considered where residual hedging risk is quantified through convex measures. The main objective of this current paper is in assessing the impact of including options within hedges on the equal risk price $C_0^{\star}$ and on the level of market incompleteness quantified by our $\epsilon^{\star}$-metrics. A specific focus is on the examination of the interplay between different stylized features of equity jump and volatility risks and the use of options as hedging instruments within the ERP framework.
	The numerical scheme of \cite{carbonneau2020equal}, which relies on the deep hedging algorithm of \cite{buehler2019deep}, is used to solve the embedded global hedging problems of the ERP framework through the representation of the long and short trading policies with two distinct long-short term memory (LSTM) neural networks. 

	Sensitivity analyses with Monte Carlo simulations are performed under several empirically plausible sets of parameters for the jump and volatility risk models in order to highlight the impact of different stylized features of the models on $C_0^{\star}$ and $\epsilon^{\star}$. Numerical values indicate that in the presence of jump risk, hedging with options entails a significant reduction of both equal risk prices and market incompleteness as compared to hedging solely with the underlying stock. The latter stems from the fact that using options as hedging instruments rather than only the underlying stock shrinks the asymmetry of tail risk, which tends to both shrink option prices and reduce market incompleteness. On the other hand, in the presence of volatility risk, while option hedges can reduce equal risk prices as compared to stock hedges, the impact can be marginal and is highly sensitive to the moneyness level of the put option being priced as well as to the maturity of traded options. This can be explained by the fact that while the impact of jump risk dynamics considered in this paper is asymmetric by entailing significantly more weight on the right (resp. left) tail of the short (resp. long) hedging error through predominantly negative jumps, volatility risk impacts both upside and downside risk. Furthermore, additional experiments conducted show that the average price level of short-term options used as hedging instruments is effectively reflected into the equal risk price of longer-term options. The latter highlights the potential of the ERP framework as a fair valuation approach providing prices consistent with observable market prices. Thus, ERP could be applied for instance in the context of pricing over-the-counter derivatives with vanilla calls and puts hedges or pricing less liquid long-term derivatives (e.g. LEAPS contracts) with shorter-term liquid options.
	
	Moreover, the benchmarking of equal risk prices to variance-optimal derivative premiums $C_0^{(VO)}$ is performed. The deep hedging algorithm is also used as the numerical scheme to solve the variance-optimal hedging problems. Numerical results show that while $C_0^{\star}$ tends to be larger than $C_0^{(VO)}$, trading options entails much smaller disparity between equal risk and variance-optimal prices as compared to trading only the underlying stock in the presence of jump or volatility risk. The latter is due to the market incompleteness being significantly smaller when option hedges are used to mitigate jump and volatility risks. Furthermore, additional experiments conducted demonstrate the ability of ERP to span a large interval of prices through the choice of convex risk measures, which is close to encompass the variance-optimal premium.

	
	
	\section{Acknowledgements}
	Alexandre Carbonneau gratefully acknowledges financial support from the Fonds de recherche du Qu\'ebec - Nature et technologies (FRQNT, grant number 205683) and The Montreal Exchange. Fr{\'e}d{\'e}ric Godin gratefully acknowledges financial support from Natural Sciences and Engineering Research Council of Canada (NSERC, grant number RGPIN-2017-06837).
	

	\bibliographystyle{apalike}
	\bibliography{Biblio_ERP_multi_asset}


\appendix


\section{Variance-optimal hedging}
\label{appendix:QDH_price}
Denote $J^{(VO)}: \mathbb{R}^{q} \times \mathbb{R} \rightarrow \mathbb{R}$ as the cost function to be minimized for variance-optimal procedures:
\begin{align}
J^{(VO)}(\theta, V_0) :=\E \left[\left(\Phi(S_{N}^{(0,b)}) -B_{N}(V_0 + G_{N}^{\delta^{\theta}})\right)^{2}\right], \quad (\theta, V_0) \in \mathbb{R}^{q} \times \mathbb{R}, \label{eq:ref_QDH_cost_func}
\end{align}	
where $\theta$ is the set of trainable parameters of the LSTM $F_{\theta}$, $V_0$ is the initial capital investment and $\delta^{\theta}$ is to be understood as the output sequence of $F_{\theta}$. Let $\tilde{\theta}:=\{\theta, V_{0}\}$ be the augmented set of trainable parameters which includes the initial portfolio value. Minibatch SGD with Monte Carlo sampling can naturally also be used to minimize \eqref{eq:ref_QDH_cost_func} jointly for the trainable parameters and the initial capital investment by updating iteratively the augmented set $\tilde{\theta}$:
\begin{align}
\tilde{\theta}_{j+1} &= \tilde{\theta}_{j} - \eta_j \nabla_{\tilde{\theta}}\hat{J}^{(VO)}(\mathbb{B}_j, V_{0,j}), \label{eq:ref_SGD_iterative_param_approx_QDH}
\end{align}
where $\tilde{\theta}_0:=\{\theta_0, V_{0,0}\}$ is the initial set\footnote{
	As described in \cref{subsec:dynamics_IV}, an implied volatility dynamics is considered to price options used as hedging instruments. In numerical experiments of \cref{section:numerical_results}, $V_{0,0}$ is set at the price obtained with the time-$0$ implied volatility. The authors also tested the naive initialization scheme $V_{0,0} = 0$ as a robustness test, and found that the resulting variance-optimal premiums were marginally affected by this choice. Also, the Glorot uniform initialization of \cite{glorot2010understanding} is used to select $\theta_0$.
} and $\hat{J}^{(VO)}(\mathbb{B}_j, V_{0,j})$ is the empirical estimator of 
$J^{(VO)}(\theta, V_0)$
evaluated with the minibatch of hedging errors $\mathbb{B}_j = \{\Phi(S_{N,i}^{(0,b)}) -B_{N}(V_{0,j} + G_{N,i}^{\delta^{\theta_j}})\}_{i=1}^{N_{\text{\text{batch}}}}$ when $\tilde{\theta} = ´\tilde{\theta}_j$ (i.e. $\theta = \theta_j$ and $V_0 = V_{0,j}$):
\begin{align}
\hat{J}^{(VO)}(\mathbb{B}_j, V_{0,j}):= \frac{1}{N_{\text{batch}}}\sum_{i=1}^{N_{\text{batch}}}\left(\Phi(S_{N,i}^{(0,b)}) -B_{N}(V_{0,j} + G_{N,i}^{\delta^{\theta_j}})\right)^{2}. \label{eq:ref_cost_func_quadratic}
\end{align}

	

\section{Pseudo-code deep hedging}
\label{appendix:pseudo_code}
\cref{algo:pseudo_code} presents the pseudo-code to perform a one-step update of the trainable parameters as in \eqref{eq:ref_SGD_iterative_param_approx} for the global hedging problems of the ERP framework, i.e. updating $\theta_j$ to $\theta_{j+1}$. For convenience, the pseudo-code is presented for the case of trading exclusively the underlying stock and for the short position trading policy, but it is trivial to generalize to the case of trading other hedging instruments (e.g. short-term options) and for the long position trading policy. Note that the pseudo-code is described for the MJD dynamics, but it can be generalized to the GARCH dynamics by sampling log-returns from \eqref{eq:ref_GARCH} in line $(6)$, and adding the stochastic volatilities to feature vectors as described in \cref{subsubsec:GARCH_model}. Furthermore, the pseudo-code can also easily be extended to variance-optimal hedging by updating the augmented set $\tilde{\theta}_{j}$ to $\tilde{\theta}_{j+1}$ with \eqref{eq:ref_SGD_iterative_param_approx_QDH} instead of $\theta_j$ to $\theta_{j+1}$ in line $(17)$ and by adapting the empirical cost function in line $(15)$ to \eqref{eq:ref_cost_func_quadratic}. Lastly, recall that a GitHub repository with samples of codes in Python for the training procedure of neural networks is available online: \href{https://github.com/alexandrecarbonneau}{github.com/alexandrecarbonneau}. The implementation replicates results of  \cref{table:ERP_one_year_put_MJD_analysis} with jump risk scenario $2$, and can easily be adapted to reproduce all results presented in \cref{section:numerical_results}.

\begin{algorithm}
	\caption{Pseudo-code short trading policy with stock hedges under the MJD model\\
		Input: $\theta_j$ \\
		Output: $\theta_{j+1}$}
	\label{algo:pseudo_code}
	\begin{algorithmic}[1]
		\For {$i=1,\ldots,N_{\text{batch}}$}  \Comment{Loop over each path of minibatch}
		\State $X_{0,i} =[\log(S_{0,i}^{(0,b)}/K), V_{0,i}^{\delta}]$ \Comment{Time-$0$ feature vector of $F_{\theta}^{(\mathcal{S})}$ with $V_{0,i}^{\delta}=0$}
		\For {$n=0,\ldots,N-1$} 
		\State $Y_{n,i} \leftarrow$ \text{time-$t_{n}$ output of LSTM $F_{\theta}^{(\mathcal{S})}$ with $\theta = \theta_j$} 		
		\State $\delta_{n+1,i}^{(0)} = Y_{n,i}$		  
		\State $y_{n+1,i} \sim$ \eqref{eq:ref_MJD}  \Comment{Sample next log-return}
		\State $S_{n+1,i}^{(0,b)} = S_{n,i}^{(0,b)} e^{y_{n+1,i}}$
		\State $V_{n+1,i}^{\delta} = e^{r\Delta_N}V_{n,i}^{\delta} + \delta_{n+1,i}^{(0)}(S_{n+1,i}^{(0,b)} - e^{r\Delta_N} S_{n,i}^{(0,b)})$ \Comment{See \eqref{eq:ref_update_rule_port_value} for details}
		\State $X_{n+1,i} =[\log(S_{n+1,i}^{(0,b)}/K), V_{n+1,i}^{\delta}]$		 \Comment{
			Time-$t_{n+1}$ feature vector for $F_{\theta}^{\mathcal{(S)}}$}
		\EndFor
		\State $\Phi(S_{N,i}^{(0,b)})=\max(K-S_{N,i}^{(0,b)}, 0)$
		\State $\pi_{i,j}=\Phi(S_{N,i}^{(0,b)}) - V_{N,i}^{\delta}$
		\EndFor
		\State $\reallywidehat{\text{VaR}}_{\alpha} = \pi_{[\tilde{N}], j}$ \Comment{$\tilde{N}$$^{\text{th}}$ ordered hedging error with $\tilde{N}:= \ceil*{\alpha N_{\text{batch}}}$}
		\State $\reallywidehat{\text{CVaR}}_{\alpha}= \reallywidehat{\text{VaR}}_{\alpha} + \frac{1}{(1-\alpha)N_{\text{batch}}}\sum_{i=1}^{N_{\text{batch}}}\max(\pi_{i,j}-\reallywidehat{\text{VaR}}_{\alpha},0)$
		\State $\eta_j \leftarrow$ Adam algorithm
		\State $\theta_{j+1} = \theta_{j} - \eta_{j} \nabla_{\theta} \reallywidehat{\text{CVaR}}_{\alpha}$ \Comment{$\nabla_{\theta} \reallywidehat{\text{CVaR}}_{\alpha}$ computed with Tensorflow}
	\end{algorithmic}
	Notes: Subscript $i$ represents the $i^{th}$ simulated path among the minibatch of size $N_{\text{batch}}.$ Also, the time-$0$ feature vector is fixed for all paths, i.e. $S_{0,i}^{(0,b)} = S_{0}^{(0,b)}$ and $V_{0,i}^{\delta}=V_{0}^{\delta}=0$.
\end{algorithm}

	\end{document}


\title{\Large \bfseries Supplementary Material:  Deep Equal Risk Pricing of Financial Derivatives with Multiple Hedging Instruments} 
	
	\author[a] {Alexandre Carbonneau\thanks{Corresponding author\vspace{0.2em} \newline
	{\it Email addresses:} \href{mailto:alexandre.carbonneau@mail.concordia.ca}{alexandre.carbonneau@mail.concordia.ca} (Alexandre Carbonneau), \href{mailto:frederic.godin@concordia.ca}{frederic.godin@concordia.ca} (Fr\'ed\'eric Godin).}}
	\author[b]{Fr\'ed\'eric Godin}
	\affil[a,b]{{\small Concordia University, Department of Mathematics and Statistics, Montr\'eal, Canada}}
	
	\vspace{-10pt}
	\date{ 
		\today}
	
	\maketitle \thispagestyle{empty} 

	This supplementary material presents additional numerical experiments of the ERP framework. 
	\cref{sec:BSM_ERP_analysis} presents values for $C_0^{\star}$ and $\epsilon^{\star}$ under the Black-Scholes model with the same setup as in Table $3$ and Table $5$. 
	\cref{sec:sens_avg_IV} presents the sensitivity analysis of $C_0^{\star}$ and $\epsilon^{\star}$ to implied volatility risk with a similar setup as in Table $5$.
	\cref{sec:add_sens_analysis_convex} presents the corresponding sensitivity analysis of Table $6$ under jump risk scenario $1$ and $3$ as well as under the three sets of parameters for volatility risk. 
	\cref{sec:add_benchmark_ERP_VO} presents the benchmarking of $C_0^{\star}$ to $C_0^{(VO)}$ under the BSM and GARCH dynamics with the same setup as in Table $7$ and Table $8$.

	\section{Sensitivity of equal risk pricing under Black-Scholes}
	\label{sec:BSM_ERP_analysis}
	The Black-Scholes dynamics models log-returns as independent and identically distributed Gaussian random variables of daily periodic mean $\frac{1}{252}(\mu - \frac{\sigma^{2}}{2})$ and variance $\frac{\sigma^{2}}{252}$. Daily log-returns under this model have the representation
	\begin{align}
	\tilde{y}_{j,n} = \frac{1}{252}\left(\mu - \frac{\sigma^{2}}{2}\right) + \sigma \sqrt{\frac{1}{252}}\epsilon_{j,n}, \label{eq:ref_BSM_log_ret}
	\end{align}
	where $\mu$ and $\sigma$ are the parameters on a yearly scale. Under the BSM, feature vectors of the neural networks do not require the inclusion of additional state variables (i.e. $\varphi_n = 0$ for all time steps $n$ in $(3.4)$ and $(3.5)$). 
	\cref{table:ERP_one_year_put_BSM_analysis} presents the equal risk prices $C_0^{\star}$ and $\epsilon^{\star}$-metrics
	under the BSM with $\mu=0.1$ and $\sigma = 0.10, 0.15$ and $0.20$.
	
	\begin{table}
		\caption {Equal risk prices $C_{0}^{\star}$ and residual hedging risk $\epsilon^{\star}$ under the Black-Scholes model for OTM ($K=90$), ATM ($K=100$) and ITM ($K=110$) put options of maturity $T=1$.} \label{table:ERP_one_year_put_BSM_analysis}
		\renewcommand{\arraystretch}{1.15}
		\begin{adjustwidth}{-1in}{-1in} 
			\centering
			\begin{tabular}{lccccccccccc}
				\hline\noalign{\smallskip}
				& \multicolumn{3}{c}{$\text{OTM}$} & & \multicolumn{3}{c}{$\text{ATM}$} & & \multicolumn{3}{c}{$\text{ITM}$} \\
				\cline{2-4}\cline{6-8}\cline{10-12} Annual volatility ($\sigma$)  & $10\%$ & $15\%$ & $20\%$ &   & $10\%$ & $15\%$ & $20\%$ &   & $10\%$ & $15\%$ & $20\%$ \\
				\hline\noalign{\medskip} 
				\multicolumn{4}{l}{\underline{\emph{$C_{0}^{\star}$}}} \\\noalign{\smallskip} 
				$\text{Daily stock}$        & $0.43$   & $1.48$     & $2.83$   &  & $2.65$   & $4.60$      & $6.68$    & & $8.37$   & $10.19$     & $12.08$  \\
				$\text{Monthly stock}$      & $0.49$   & $1.67$     & $3.15$   &  & $2.71$   & $4.70$      & $6.72$    & & $8.56$   & $10.42$     & $12.39$  \\
				$\text{1-month options}$    & $0.44$   & $1.56$     & $3.00$   &  & $2.74$   & $4.70$      & $6.71$    & & $8.48$   & $10.30$     & $12.27$  \\
				$\text{3-months options}$   & $0.44$   & $1.59$     & $3.08$   &  & $2.81$   & $4.80$      & $6.82$    & & $8.58$   & $10.47$     & $12.51$  \\
				&   &   &  & & & & & & & & \\
				\multicolumn{4}{l}{\underline{\emph{$\epsilon^{\star}$}}} \\\noalign{\smallskip} 
				$\text{Daily stock}$        & $0.19$   & $0.41$     & $0.65$   &  & $0.38$   & $0.64$      & $1.00$    & & $0.39$   & $0.67$     & $0.88$  \\
				$\text{Monthly stock}$      & $0.50$   & $1.52$     & $2.59$   &  & $1.54$   & $2.53$      & $3.48$    & & $1.51$   & $2.59$     & $3.60$  \\
				$\text{1-month options}$    & $0.19$   & $0.57$     & $0.94$   &  & $0.57$   & $0.91$      & $1.30$    & & $0.64$   & $1.01$     & $1.37$  \\
				$\text{3-months options}$   & $0.29$   & $0.93$     & $1.55$   &  & $0.96$   & $1.55$      & $2.11$    & & $1.09$   & $1.76$     & $2.35$  \\
				&   &   &  & & & & & & & & \\
				\multicolumn{4}{l}{\underline{\emph{$\epsilon^{\star}/C_{0}^{\star}$}}} \\\noalign{\smallskip} 
				$\text{Daily stock}$        & $0.44$   & $0.28$     & $0.23$   &  & $0.14$   & $0.14$      & $0.15$    & & $0.05$   & $0.07$     & $0.07$  \\
				$\text{Monthly stock}$      & $1.01$   & $0.91$     & $0.82$   &  & $0.57$   & $0.54$      & $0.52$    & & $0.18$   & $0.25$     & $0.29$  \\
				$\text{1-month options}$    & $0.44$   & $0.37$     & $0.31$   &  & $0.21$   & $0.19$      & $0.19$    & & $0.07$   & $0.10$     & $0.11$  \\
				$\text{3-months options}$   & $0.67$   & $0.59$     & $0.50$   &  & $0.34$   & $0.32$      & $0.31$    & & $0.13$   & $0.17$     & $0.19$  \\
				\noalign{\medskip}\hline
			\end{tabular}%
		\end{adjustwidth}
		Notes: Results are computed based on $100,\!000$ independent paths generated from the Black-Scholes model for the underlying with $\mu=0.1$ and $\sigma=0.1, 0.15$ and $0.20$ (see \eqref{eq:ref_BSM_log_ret} for model description). \textit{Hedging instruments}: daily or monthly rebalancing with the underlying stock and 1-month or 3-months options with ATM calls and puts. Options used as hedging instruments are priced with implied volatility modeled as a log-AR(1) dynamics with $\kappa = 0.15, \sigma_{\text{IV}} = 0.06$ and $\varrho = -0.6$ for all cases, and $\vartheta$ set to $\log(0.10), \log(0.15)$ and $\log(0.20)$ when $\sigma = 0.10, 0.15$ and $0.20$, respectively (see Section $4.1.4$ for log-AR(1) model description). The training of neural networks is done as described in Section $4.1.5$. The confidence level of the CVaR measure is $\alpha = 0.95$.
	\end{table}

	\section{Sensitivity of equal risk pricing to implied volatility}
	\label{sec:sens_avg_IV}
	\cref{table:ERP_one_year_put_IV_analysis} presents the sensitivity analysis of the ERP framework to the long-run implied volatility when trading $1$-month or $3$-months options. These options are priced with implied volatility modeled as a log-AR(1) dynamics with parameters of Table $2$ for $\{\kappa, \sigma_{\text{IV}}, \varrho\}$ (i.e. $\kappa = 0.15, \sigma_{\text{IV}} = 0.06$ and $\varrho = -0.6$), and the long-run parameter $\vartheta$ set to $\log(0.14), \log(0.15)$ and $\log(0.16)$, respectively. The underlying stock dynamics is the GARCH model with the set of parameters corresponding to the stationary yearly volatility of $15\%$ in Table $4$.
	
	\begin{table}
		\caption {Sensitivity analysis of equal risk prices $C_{0}^{\star}$ and residual hedging risk $\epsilon^{\star}$ to implied volatility (IV) risk for OTM ($K=90$), ATM ($K=100$) and ITM ($K=110$) put options of maturity $T=1$.} \label{table:ERP_one_year_put_IV_analysis}
		\renewcommand{\arraystretch}{1.15}
		\begin{adjustwidth}{-1in}{-1in} 
			\centering
			\begin{tabular}{lccccccccccc}
				\hline\noalign{\smallskip}
				& \multicolumn{3}{c}{$\text{OTM}$} & & \multicolumn{3}{c}{$\text{ATM}$} & & \multicolumn{3}{c}{$\text{ITM}$} \\
				\cline{2-4}\cline{6-8}\cline{10-12} Long-run IV   & $14\%$ & $15\%$ & $16\%$ &   & $14\%$ & $15\%$ & $16\%$ &   & $14\%$ & $15\%$ & $16\%$ \\
				\hline\noalign{\medskip} 
				\multicolumn{4}{l}{\underline{\emph{$C_{0}^{\star}$}}} \\\noalign{\smallskip} 
				$\text{1-month options}$    & $1.52$   & $1.74$     & $1.98$   &  & $4.52$   & $4.87$      & $5.20$    & & $10.01$   & $10.32$     & $10.66$  \\
				$\text{3-months options}$   & $1.86$   & $2.07$     & $2.28$   &  & $4.75$   & $5.08$      & $5.39$    & & $10.08$   & $10.38$     & $10.66$  \\
				&   &   &  & & & & & & & & \\
				\multicolumn{4}{l}{\underline{\emph{$\epsilon^{\star}$}}} \\\noalign{\smallskip} 
				$\text{1-month options}$    & $0.61$   & $0.65$     & $0.66$   &  & $0.96$   & $1.00$      & $0.96$    & & $0.99$   & $1.03$     & $1.01$  \\
				$\text{3-months options}$   & $1.28$   & $1.32$     & $1.36$   &  & $1.72$   & $1.77$      & $1.78$    & & $1.63$   & $1.68$     & $1.72$  \\
				&   &   &  & & & & & & & & \\
				\multicolumn{4}{l}{\underline{\emph{$\epsilon^{\star}/C_{0}^{\star}$}}} \\\noalign{\smallskip} 
				$\text{1-month options}$    & $0.40$   & $0.37$     & $0.34$   &  & $0.21$   & $0.20$      & $0.18$    & & $0.10$   & $0.10$     & $0.09$  \\
				$\text{3-months options}$   & $0.69$   & $0.64$     & $0.60$   &  & $0.36$   & $0.35$      & $0.33$    & & $0.16$   & $0.16$     & $0.16$  \\
				\noalign{\medskip}\hline
			\end{tabular}%
		\end{adjustwidth}
		Notes: Results are computed based on $100,\!000$ independent paths generated from the GJR-GARCH(1,1) model for the underlying with $\mu=3.968\text{e-}04,$ $\omega=1.964\text{e-}06$, $\upsilon = 0.05$, $\gamma = 0.6$ and $\beta = 0.91$ which entails stationary yearly volatility of $15\%$ (see Section $4.1.3$ for model description). Hedging instruments consist of 1-month or 3-months options with ATM calls and puts. The latter options are priced with implied volatility modeled as a log-AR(1) dynamics with $\kappa = 0.15, \sigma_{\text{IV}} = 0.06$ and $\varrho = -0.6$ for all cases, and the long-run parameter $\vartheta$ set to $\log(0.14), \log(0.15)$ or $\log(0.16)$ (see Section $4.1.4$ for the log-AR(1) model description). The training of neural networks is done as described in Section $4.1.5$. The confidence level of the CVaR measure is $\alpha = 0.95$.
	\end{table}
		
	\section{Additional sensitivity analysis convex risk measure}
	\label{sec:add_sens_analysis_convex}
	\cref{table:sensitivity_analysis_CVAR_090_095_099_jump_risk} presents the sensitivity analysis of the ERP framework to the choice of convex risk measure when trading $3$-months options under jump risk scenario $1$ and $3$ of Table $1$. \cref{table:sensitivity_analysis_CVAR_090_095_099_vol_risk} presents this sensitivity analysis under volatility risk with the three sets of parameters described in Table $4$.
	\begin{table} [ht]
		\caption {Sensitivity analysis of equal risk prices $C_{0}^{\star}$ and residual hedging risk $\epsilon^{\star}$ for OTM ($K=90$), ATM ($K=100$) and ITM ($K=110$) put options of maturity $T=1$ under the MJD dynamics with jump risk scenario $1$ and scenario $3$.} \label{table:sensitivity_analysis_CVAR_090_095_099_jump_risk}
		\renewcommand{\arraystretch}{1.15}
		\begin{adjustwidth}{-1in}{-1in} 
			\centering
			\begin{tabular}{lccccccccccc}
				\hline\noalign{\smallskip}
				& \multicolumn{3}{c}{$C_{0}^{\star}$} & & \multicolumn{3}{c}{$\epsilon^{\star}$} & & \multicolumn{3}{c}{$\epsilon^{\star}/C_{0}^{\star}$} \\
				\cline{2-4}\cline{6-8}\cline{10-12}   Moneyness   & $\text{OTM}$ & $\text{ATM}$ & $\text{ITM}$ &   & $\text{OTM}$ & $\text{ATM}$ & $\text{ITM}$ &   & $\text{OTM}$ & $\text{ATM}$ & $\text{ITM}$ \\
				\hline\noalign{\medskip} 
				%
				\multicolumn{4}{l}{\underline{\textbf{{\emph{Jump risk scenario $1$}}}}} \\\noalign{\smallskip} 
				$\text{CVaR}_{0.90}$ & $1.63$   & $4.76$  & $10.32$  &   & $0.78$    & $1.28$  & $1.42$  &   & $0.48$  & $0.27$  & $0.14$ \\
				$\text{CVaR}_{0.95}$ & $6\%$    & $2\%$   & $1\%$    &   & $32\%$    & $24\%$  & $20\%$  &   & $24\%$  & $21\%$  & $19\%$ \\
				$\text{CVaR}_{0.99}$ & $28\%$   & $7\%$   & $4\%$    &   & $103\%$   & $70\%$  & $65\%$  &   & $59\%$  & $58\%$  & $58\%$ \\
				\\
				\multicolumn{4}{l}{\underline{\textbf{{\emph{Jump risk scenario $3$}}}}} \\\noalign{\smallskip} 
				$\text{CVaR}_{0.90}$ & $2.07$   & $5.08$   & $10.44$  &   & $1.21$    & $1.61$   & $1.57$   &   & $0.58$  & $0.32$  & $0.15$ \\
				$\text{CVaR}_{0.95}$ & $15\%$   & $4\%$    & $1\%$    &   & $39\%$    & $25\%$   & $20\%$   &   & $21\%$  & $21\%$  & $19\%$ \\
				$\text{CVaR}_{0.99}$ & $56\%$   & $13\%$   & $5\%$    &   & $128\%$   & $76\%$   & $68\%$   &   & $47\%$  & $56\%$  & $60\%$ \\
				\noalign{\medskip}\hline
			\end{tabular}%
		\end{adjustwidth}
		Notes: Results are computed based on $100,\!000$ independent paths generated from the Merton Jump-Diffusion model for the underlying (see Section $4.1.2$ for model description). Parameters for jump risk scenario $1$ and $3$ are presented in Table $1$.
		Hedging instruments consist of 3-months ATM calls and puts priced with implied volatility modeled with a log-AR(1) dynamics (see Section $4.1.4$ for model description and Table $2$ for parameters values). The training of neural networks is done as described in Section $4.1.5$.  Values for the $\text{CVaR}_{0.95}$ and $\text{CVaR}_{0.99}$ measures are expressed relative to $\text{CVaR}_{0.90}$ (\% increase).
	\end{table}
	\begin{table} [ht]
	\caption {Sensitivity analysis of equal risk prices $C_{0}^{\star}$ and residual hedging risk $\epsilon^{\star}$ for OTM ($K=90$), ATM ($K=100$) and ITM ($K=110$) put options of maturity $T=1$ under the GARCH dynamics.} \label{table:sensitivity_analysis_CVAR_090_095_099_vol_risk}
	\renewcommand{\arraystretch}{1.15}
	\begin{adjustwidth}{-1in}{-1in} 
		\centering
		\begin{tabular}{lccccccccccc}
			\hline\noalign{\smallskip}
			& \multicolumn{3}{c}{$C_{0}^{\star}$} & & \multicolumn{3}{c}{$\epsilon^{\star}$} & & \multicolumn{3}{c}{$\epsilon^{\star}/C_{0}^{\star}$} \\
			\cline{2-4}\cline{6-8}\cline{10-12}   Moneyness   & $\text{OTM}$ & $\text{ATM}$ & $\text{ITM}$ &   & $\text{OTM}$ & $\text{ATM}$ & $\text{ITM}$ &   & $\text{OTM}$ & $\text{ATM}$ & $\text{ITM}$ \\
			\hline\noalign{\medskip} 
			%
			\multicolumn{4}{l}{\underline{\textbf{{\emph{Stationary volatility: $10\%$}}}}} \\\noalign{\smallskip} 
			$\text{CVaR}_{0.90}$ & $0.65$   & $2.89$   & $8.42$   &   & $0.40$    & $0.83$  & $0.86$  &   & $0.61$  & $0.29$  & $0.10$ \\
			$\text{CVaR}_{0.95}$ & $17\%$   & $4\%$    & $1\%$    &   & $46\%$    & $32\%$  & $21\%$  &   & $25\%$  & $27\%$  & $20\%$ \\
			$\text{CVaR}_{0.99}$ & $70\%$   & $13\%$   & $3\%$    &   & $159\%$   & $94\%$  & $66\%$  &   & $52\%$  & $72\%$  & $60\%$ \\
			\\
			\multicolumn{4}{l}{\underline{\textbf{{\emph{Stationary volatility: $15\%$}}}}} \\\noalign{\smallskip} 
			$\text{CVaR}_{0.90}$ & $1.88$  & $4.91$   & $10.28$  &   & $0.97$    & $1.39$  & $1.40$  &   & $0.51$  & $0.28$  & $0.14$ \\
			$\text{CVaR}_{0.95}$ & $10\%$  & $4\%$    & $1\%$    &   & $36\%$    & $27\%$  & $21\%$  &   & $24\%$  & $23\%$  & $19\%$ \\
			$\text{CVaR}_{0.99}$ & $38\%$  & $11\%$   & $4\%$    &   & $122\%$   & $81\%$  & $69\%$  &   & $61\%$  & $63\%$  & $63\%$ \\
			\\
			\multicolumn{4}{l}{\underline{\textbf{{\emph{Stationary volatility: $20\%$}}}}} \\\noalign{\smallskip} 
			$\text{CVaR}_{0.90}$ & $3.40$   & $6.95$   & $12.27$  &   & $1.54$    & $1.93$  & $1.89$  &   & $0.45$  & $0.28$  & $0.15$ \\
			$\text{CVaR}_{0.95}$ & $7\%$    & $3\%$    & $1\%$    &   & $31\%$    & $25\%$  & $22\%$  &   & $22\%$  & $21\%$  & $20\%$ \\
			$\text{CVaR}_{0.99}$ & $28\%$   & $10\%$   & $5\%$    &   & $103\%$   & $77\%$  & $71\%$  &   & $58\%$  & $61\%$  & $64\%$ \\
			\noalign{\medskip}\hline
		\end{tabular}%
	\end{adjustwidth}
	Notes: Results are computed based on $100,\!000$ independent paths generated from the GJR-GARCH(1,1) model for the underlying with three sets of parameters implying stationary yearly volatilities of $10\%, 15\%$ and $20\%$ (see Section $4.1.3$ for model description and Table $4$ for parameters values). Hedging instruments consist of 3-months ATM calls and puts priced with implied volatility modeled with a log-AR(1) dynamics with $\kappa = 0.15, \sigma_{\text{IV}} = 0.06$ and $\varrho = -0.6$ for all cases, and $\vartheta$ set to $\log(0.10), \log(0.15)$ and $\log(0.20)$ when the GARCH stationary volatility is $10\%, 15\%$ and $20\%$, respectively (see Section $4.1.4$ for log-AR(1) model description).
	The training of neural networks is done as described in Section $4.1.5$. Values for the $\text{CVaR}_{0.95}$ and $\text{CVaR}_{0.99}$ measures are expressed relative to $\text{CVaR}_{0.90}$ (\% increase).
	\end{table}

	\section{Benchmarking of equal risk prices to variance-optimal premiums}
	\label{sec:add_benchmark_ERP_VO} 
	\cref{table:ERP_RN_prices_one_year_BSM} and \cref{table:ERP_RN_prices_one_year_VOL} present the benchmarking of $C_0^{\star}$ with the $\text{CVaR}_{0.95}$ risk measure to $C_0^{(VO)}$ with the same setup as Table $7$ under the Black-Scholes and GARCH dynamics, respectively. Furthermore, \cref{table:ERP_vs_VO_sensitivity_alpha_BSM} and \cref{table:ERP_vs_VO_sensitivity_alpha_VOL} perform the benchmarking of $C_0^{\star}$ to $C_0^{(VO)}$ with the $\text{CVaR}_{0.90}, \text{CVaR}_{0.95}$ and $\text{CVaR}_{0.99}$ measures and $3$-months option hedges under the BSM and GARCH dynamics, respectively. The latter experiments are performed with the same setup as Table $8$ under the MJD dynamics.
	%
	\begin{table}[ht]
	\caption {Equal risk prices $C_{0}^{\star}$ and variance-optimal (VO) prices $C_0^{(VO)}$ under the Black-Scholes model for OTM ($K=90$), ATM ($K=100$) and ITM ($K=110$) put options of maturity $T=1$.} \label{table:ERP_RN_prices_one_year_BSM}
	\renewcommand{\arraystretch}{1.15}
	\begin{adjustwidth}{-1in}{-1in} 
		\centering
		\begin{tabular}{lccccccccccc}
			\hline\noalign{\smallskip}
			& \multicolumn{3}{c}{$\text{OTM}$} & & \multicolumn{3}{c}{$\text{ATM}$} & & \multicolumn{3}{c}{$\text{ITM}$} \\
			\cline{2-4}\cline{6-8}\cline{10-12}   $\text{Annual volatility } (\sigma)$   & $10\%$ & $15\%$ & $20\%$ &   & $10\%$ & $15\%$ & $20\%$ &   & $10\%$ & $15\%$ & $20\%$ \\
			\hline\noalign{\medskip} 
			\multicolumn{4}{l}{\underline{\emph{$C_0^{(VO)}$}}} \\\noalign{\smallskip} 
			$\text{Daily stock}$        & $0.38$   & $1.39$     & $2.77$    &  & $2.62$   & $4.53$     & $6.46$    & & $8.34$   & $10.12$       & $12.04$  \\
			$\text{Monthly stock}$      & $0.35$   & $1.36$     & $2.73$    &  & $2.55$   & $4.47$     & $6.39$    & & $8.27$   & $10.07$       & $11.98$  \\
			$\text{1-month options}$    & $0.42$   & $1.47$     & $2.84$    &  & $2.67$   & $4.58$     & $6.52$    & & $8.35$   & $10.13$       & $12.06$  \\
			$\text{3-months options}$   & $0.39$   & $1.44$     & $2.84$    &  & $2.65$   & $4.56$     & $6.50$    & & $8.32$   & $10.12$       & $12.04$  \\
			&   &   &  & & & & & & & & \\
			\multicolumn{4}{l}{\underline{\emph{$C_{0}^{\star}$}}} \\\noalign{\smallskip} 
			$\text{Daily stock}$        & $12\%$   & $6\%$      & $2\%$    &  & $1\%$   & $2\%$    & $3\%$    & & $0\%$   & $1\%$     & $0\%$  \\
			$\text{Monthly stock}$      & $40\%$   & $23\%$     & $15\%$   &  & $6\%$   & $5\%$    & $5\%$    & & $3\%$   & $4\%$     & $3\%$  \\
			$\text{1-month options}$    & $5\%$    & $6\%$      & $6\%$    &  & $3\%$   & $3\%$    & $3\%$    & & $2\%$   & $2\%$     & $2\%$  \\
			$\text{3-months options}$   & $11\%$   & $10\%$     & $9\%$   &  & $6\%$   & $5\%$    & $5\%$    & & $3\%$   & $3\%$     & $4\%$  \\
			\noalign{\medskip}\hline
		\end{tabular}%
	\end{adjustwidth}
	Notes: Results are computed based on $100,\!000$ independent paths generated from the Black-Scholes model for the underlying with $\mu=0.1$ and $\sigma=0.1, 0.15$ and $0.20$ (see \eqref{eq:ref_BSM_log_ret} for model description). \textit{Hedging instruments}: daily or monthly rebalancing with the underlying stock and 1-month or 3-months options with ATM calls and puts. Options used as hedging instruments are priced with implied volatility modeled as a log-AR(1) dynamics with $\kappa = 0.15, \sigma_{\text{IV}} = 0.06$ and $\varrho = -0.6$ for all cases, and $\vartheta$ set to $\log(0.10), \log(0.15)$ and $\log(0.20)$ when $\sigma = 0.10, 0.15$ and $0.20$, respectively (see Section $4.1.4$ for log-AR(1) model description). The training of neural networks for ERP and VO hedging is done as described in Section $4.1.5$ and Appendix A, respectively. The confidence level of the CVaR measure is $\alpha = 0.95$. $C_0^{\star}$ are expressed relative to $C_0^{(VO)}$ (\% increase).
	\end{table}


\begin{table}[ht]
	\caption {Equal risk prices $C_{0}^{\star}$ and variance-optimal (VO) prices $C_0^{(VO)}$ with volatility risk for OTM ($K=90$), ATM ($K=100$) and ITM ($K=110$) put options of maturity $T=1$.} \label{table:ERP_RN_prices_one_year_VOL}
	\renewcommand{\arraystretch}{1.15}
	\begin{adjustwidth}{-1in}{-1in} 
		\centering
		\begin{tabular}{lccccccccccc}
			\hline\noalign{\smallskip}
			& \multicolumn{3}{c}{$\text{OTM}$} & & \multicolumn{3}{c}{$\text{ATM}$} & & \multicolumn{3}{c}{$\text{ITM}$} \\
			\cline{2-4}\cline{6-8}\cline{10-12}   Stationary volatility   & $10\%$ & $15\%$ & $20\%$ &   & $10\%$ & $15\%$ & $20\%$ &   & $10\%$ & $15\%$ & $20\%$ \\
			\hline\noalign{\medskip} 
			\multicolumn{4}{l}{\underline{\emph{$C_0^{(VO)}$}}} \\\noalign{\smallskip} 
			$\text{Daily stock}$        & $0.78$   & $1.88$     & $3.24$    &  & $2.94$   & $4.75$     & $6.60$    & & $8.29$   & $9.98$       & $11.81$  \\
			$\text{Monthly stock}$      & $0.81$   & $1.91$     & $3.27$    &  & $2.97$   & $4.80$     & $6.62$    & & $8.30$   & $10.01$      & $11.83$  \\
			$\text{1-month options}$    & $0.50$   & $1.60$     & $3.01$   &  & $2.75$   & $4.66$     & $6.60$    & & $8.32$   & $10.10$      & $12.04$  \\
			$\text{3-months options}$   & $0.61$   & $1.74$     & $3.19$   &  & $2.79$   & $4.72$     & $6.68$    & & $8.23$   & $10.05$      & $11.99$  \\
			&   &   &  & & & & & & & & \\
			\multicolumn{4}{l}{\underline{\emph{$C_{0}^{\star}$}}} \\\noalign{\smallskip} 
			$\text{Daily stock}$        & $29\%$   & $25\%$     & $19\%$   &  & $10\%$   & $13\%$    & $10\%$    & & $3\%$   & $6\%$     & $6\%$  \\
			$\text{Monthly stock}$      & $45\%$   & $39\%$     & $30\%$   &  & $13\%$   & $13\%$    & $14\%$    & & $7\%$   & $8\%$     & $9\%$  \\
			$\text{1-month options}$    & $12\%$   & $9\%$      & $9\%$    &  & $4\%$    & $4\%$     & $4\%$     & & $2\%$   & $2\%$     & $3\%$  \\
			$\text{3-months options}$   & $24\%$   & $19\%$     & $14\%$   &  & $8\%$    & $8\%$     & $7\%$     & & $3\%$   & $3\%$     & $4\%$  \\
			\noalign{\medskip}\hline
		\end{tabular}%
	\end{adjustwidth}
	Notes: Results are computed based on $100,\!000$ independent paths generated from the GJR-GARCH(1,1) model for the underlying with three sets of parameters implying stationary yearly volatilities of $10\%, 15\%$ and $20\%$ (see Section $4.1.3$ for model description and Table $4$ for parameters values). \textit{Hedging instruments}: daily or monthly rebalancing with the underlying stock and 1-month or 3-months options with ATM calls and puts. Options used as hedging instruments are priced with implied volatility modeled as a log-AR(1) dynamics with $\kappa = 0.15, \sigma_{\text{IV}} = 0.06$ and $\varrho = -0.6$ for all cases, and $\vartheta$ set to $\log(0.10), \log(0.15)$ and $\log(0.20)$ when the GARCH stationary volatility is $10\%, 15\%$ and $20\%$, respectively (see Section $4.1.4$ for log-AR(1) model description). The training of neural networks for ERP and VO hedging is done as described in Section $4.1.5$ and Appendix A, respectively. The confidence level of the CVaR measure is $\alpha = 0.95$. $C_0^{\star}$ are expressed relative to $C_0^{(VO)}$ (\% increase).
\end{table}

\begin{table}[ht]
	\caption {Sensitivity analysis of equal risk prices $C_{0}^{\star}$ with $\text{CVaR}_{0.90}, \text{CVaR}_{0.95}$ and $\text{CVaR}_{0.99}$ measures to variance-optimal (VO) prices $C_0^{(VO)}$ under the Black-Scholes model for OTM ($K=90$), ATM ($K=100$) and ITM ($K=110$) put options of maturity $T=1$.} \label{table:ERP_vs_VO_sensitivity_alpha_BSM}
	\renewcommand{\arraystretch}{1.15}
	\begin{adjustwidth}{-1in}{-1in} 
		\centering
		\begin{tabular}{lccccccccccc}
			\hline\noalign{\smallskip}
			& \multicolumn{3}{c}{$\text{OTM}$} & & \multicolumn{3}{c}{$\text{ATM}$} & & \multicolumn{3}{c}{$\text{ITM}$} \\
			\cline{2-4}\cline{6-8}\cline{10-12}  Annual volatility ($\sigma$)   & $10\%$ & $15\%$ & $20\%$ &   & $10\%$ & $15\%$ & $20\%$ &   & $10\%$ & $15\%$ & $20\%$ \\
			%
			\hline\noalign{\medskip} 
			%
			%
			$C_0^{(VO)}$  & $0.39$   & $1.44$     & $2.84$   &  & $2.65$   & $4.56$     & $6.50$    & & $8.32$   & $10.12$     & $12.04$  \\
			%
			$C_0^{\star}(\text{CVaR}_{0.90})$  & $-2\%$    & $4\%$       & $4\%$     &  & $3\%$   & $3\%$     & $3\%$    & & $3\%$   & $2\%$     & $2\%$  \\
			$C_0^{\star}(\text{CVaR}_{0.95})$  & $11\%$    & $10\%$      & $9\%$     &  & $6\%$   & $5\%$     & $5\%$    & & $3\%$   & $3\%$     & $4\%$  \\
			$C_0^{\star}(\text{CVaR}_{0.99})$  & $44\%$    & $26\%$      & $22\%$    &  & $12\%$   & $13\%$     & $9\%$    & & $6\%$   & $7\%$     & $6\%$  \\
			\noalign{\medskip}\hline
		\end{tabular}%
	\end{adjustwidth}
	Notes: Results are computed based on $100,\!000$ independent paths generated from the Black-Scholes model for the underlying with $\mu=0.1$ and $\sigma=0.1, 0.15$ and $0.20$ (see \eqref{eq:ref_BSM_log_ret} for model description). Hedging instruments consist of 3-months ATM calls and puts priced with implied volatility modeled as a log-AR(1) dynamics with $\kappa = 0.15, \sigma_{\text{IV}} = 0.06$ and $\varrho = -0.6$ for all cases, and $\vartheta$ set to $\log(0.10), \log(0.15)$ and $\log(0.20)$ when $\sigma = 0.10, 0.15$ and $0.20$, respectively (see Section $4.1.4$ for log-AR(1) model description). The training of neural networks for ERP and VO hedging is done as described in Section $4.1.5$ and Appendix A, respectively. $C_0^{\star}$ with $\text{CVaR}_{0.90}, \text{CVaR}_{0.95}$ and $\text{CVaR}_{0.99}$ are expressed relative to $C_0^{(VO)}$ (\% increase).  
\end{table}

\begin{table}[ht]
	\caption {Sensitivity analysis of equal risk prices $C_{0}^{\star}$ with $\text{CVaR}_{0.90}, \text{CVaR}_{0.95}$ and $\text{CVaR}_{0.99}$ measures to variance-optimal (VO) prices $C_0^{(VO)}$ under volatility risk for OTM ($K=90$), ATM ($K=100$) and ITM ($K=110$) put options of maturity $T=1$.} \label{table:ERP_vs_VO_sensitivity_alpha_VOL}
	\renewcommand{\arraystretch}{1.15}
	\begin{adjustwidth}{-1in}{-1in} 
		\centering
		\begin{tabular}{lccccccccccc}
			\hline\noalign{\smallskip}
			& \multicolumn{3}{c}{$\text{OTM}$} & & \multicolumn{3}{c}{$\text{ATM}$} & & \multicolumn{3}{c}{$\text{ITM}$} \\
			\cline{2-4}\cline{6-8}\cline{10-12}   Stationary volatility   & $10\%$ & $15\%$ & $20\%$ &   & $10\%$ & $15\%$ & $20\%$ &   & $10\%$ & $15\%$ & $20\%$ \\
			%
			\hline\noalign{\medskip} 
			%
			%
			$C_0^{(VO)}$  & $0.61$   & $1.74$     & $3.19$   &  & $2.79$   & $4.72$     & $6.68$    & & $8.23$   & $10.05$     & $11.99$  \\
			%
			$C_0^{\star}(\text{CVaR}_{0.90})$  & $6\%$    & $8\%$      & $6\%$    &  & $4\%$   & $4\%$     & $4\%$    & & $2\%$   & $2\%$     & $2\%$  \\
			$C_0^{\star}(\text{CVaR}_{0.95})$  & $24\%$   & $19\%$     & $14\%$   &  & $8\%$   & $8\%$     & $7\%$    & & $3\%$   & $3\%$     & $4\%$  \\
			$C_0^{\star}(\text{CVaR}_{0.99})$  & $81\%$   & $49\%$     & $37\%$   &  & $17\%$  & $16\%$    & $15\%$   & & $6\%$   & $6\%$     & $7\%$  \\
			\noalign{\medskip}\hline
		\end{tabular}%
	\end{adjustwidth}
	Notes: Results are computed based on $100,\!000$ independent paths generated from the GJR-GARCH(1,1) model for the underlying with three sets of parameters implying stationary yearly volatilities of $10\%, 15\%$ and $20\%$ (see Section $4.1.3$ for model description and Table $4$ for parameters values). Hedging instruments consist of 3-months ATM calls and puts priced with implied volatility modeled as a log-AR(1) dynamics with $\kappa = 0.15, \sigma_{\text{IV}} = 0.06$ and $\varrho = -0.6$ for all cases, and $\vartheta$ set to $\log(0.10), \log(0.15)$ and $\log(0.20)$ when the GARCH stationary volatility is $10\%, 15\%$ and $20\%$, respectively (see Section $4.1.4$ for log-AR(1) model description). The training of neural networks for ERP and VO hedging is done as described in Section $4.1.5$ and Appendix A, respectively. $C_0^{\star}$ with $\text{CVaR}_{0.90}, \text{CVaR}_{0.95}$ and $\text{CVaR}_{0.99}$ are expressed relative to $C_0^{(VO)}$ (\% increase).  
\end{table}
